\UseRawInputEncoding 
\documentclass[lettersize,journal]{IEEEtran}
\usepackage{amsmath,amsfonts}
\usepackage{color}
\usepackage{array}
\usepackage[caption=false,font=normalsize,labelfont=sf,textfont=sf]{subfig}
\usepackage{textcomp}
\usepackage{stfloats}
\usepackage{url}
\usepackage{verbatim}
\usepackage{graphicx}
\usepackage{enumerate}
\usepackage{caption}
\usepackage{cite}
\usepackage{epstopdf}         
\usepackage{flushend}
\usepackage{hyperref}
\usepackage{doi}
\usepackage{array}
\usepackage{booktabs}
\usepackage{threeparttable}
\usepackage{hyperref}
\usepackage{cleveref}
\usepackage{textcomp}
\usepackage{stfloats}
\usepackage{url}
\usepackage{verbatim}
\usepackage{graphicx}
\usepackage{cite}
\usepackage[justification=centering]{caption}
\usepackage{booktabs}  
\usepackage{threeparttable} 
\usepackage{autobreak}  
\usepackage{algpseudocode}  
\usepackage[ruled]{algorithm2e}
\usepackage{amsmath}

\usepackage{hyperref}
\usepackage{cleveref}
\usepackage{multicol}
\usepackage{cuted}
\usepackage{amsmath}
\usepackage{multicol}
\usepackage{float}
\usepackage{placeins}
\usepackage{algorithm2e}
\usepackage[utf8]{inputenc}
\usepackage[mathscr]{euscript}

\makeatletter
\renewcommand{\@makecaption}[2]{%
  \vskip\abovecaptionskip
  \sbox\@tempboxa{#1. #2}%
  \ifdim \wd\@tempboxa >\hsize
    #1. #2\par
  \else
    \global \@minipagefalse
    \hb@xt@\hsize{\hfil\box\@tempboxa\hfil}%
  \fi
  \vskip\belowcaptionskip}
\makeatother

\begin{document}
\title{Effective Finite Time Stability Control for Human-Machine Shared Vehicle Following System }

\author{ Zihan Wang, Mengran Li, Ronghui Zhang, Jing Zhao, Chuan Hu, Xiaolei Ma , and Zhijun Qiu
\thanks{This work has been submitted to the lEEE for possible publicationCopyright may be transferred without notice, after which this version mayno longer be accessible.}
\thanks{This project is jointly supported by National Natural Science Foundation of China (Nos. 52172350, 51775565), Guangdong Basic and Applied Research Foundation (Nos. 2021B1515120032, 2022B1515120072), Guangzhou Science and Technology Plan Project (Nos. 2024B01W0079, 202206030005), Nansha Key RD Program (No. 2022ZD014), Science and Technology Planning Project of Guangdong Province (No. 2023B1212060029). \textit{(Corresponding author: Ronghui Zhang.)}}
\thanks{Zihan.Wang, Mengran.Li and Ronghui.Zhang are with Guangdong Key Laboratory of Intelligent Transportation System, School of intelligent systems engineering, Sun Yat-sen University, Guangzhou 510275, China (e-mail: wangzh579@mail2.sysu.edu.cn, limr39@mail2.sysu.edu.cn, zhangrh25@mail.sysu.edu.cn).}
\thanks{Jing.Zhao is with Automotive Engineering Lab, Department of Electromechanical Engineering, University of Macau, Macau(e-mail: jzhao@um.edu.mo).}
\thanks{Chuan.Hu is with School of Mechanical Engineering, Shanghai Jiao Tong University, 800 Dongchuan Road, Shanghai 200240, China(e-mail: zhijunqiu@ualberta.ca).}
\thanks{Xiaolei.Ma is with the Key Laboratory of Intelligent Transportation Technology and System, School of Transportation Science and Engineering, Beihang University, Beijing 100191, China(e-mail: xiaolei@buaa.edu.cn).}
\thanks{Zhijun.Qiu is with Department of Civil and Environmental Engineering, University of Alberta, Edmonton, Alberta, Canada(e-mail: zhijunqiu@ualberta.ca).}}

\markboth{IEEE Transactions on Intelligent Transportation Systems}%
{}

\maketitle

\begin{abstract}

With the development of intelligent connected vehicle technology, human-machine shared control has gained popularity in vehicle following due to its effectiveness in driver assistance. However, traditional vehicle following systems struggle to maintain stability when driver reaction time fluctuates, as these variations require different levels of system intervention. To address this issue, the proposed human-machine shared vehicle following assistance system (HM-VFAS) integrates driver outputs under various states with the assistance system. The system employs an intelligent driver model that accounts for reaction time delays, simulating time-varying driver outputs. A control authority allocation strategy is designed to dynamically adjust the level of intervention based on real-time driver state assessment.  To handle instability from driver authority switching, the proposed solution includes a two-layer adaptive finite time sliding mode controller (A-FTSMC). The first layer is an integral sliding mode adaptive controller that ensures robustness by compensating for uncertainties in the driver output. The second layer is a fast non-singular terminal sliding mode controller designed to accelerate convergence for rapid stabilization. Using real driver videos as inputs, the performance of the HM-VFAS was evaluated. Results show that the proposed control strategy maintains a safe distance under time-varying driver states, with the actual acceleration error relative to the target acceleration maintained within \ensuremath{\pm0.5\!\ \text{m/s}^2} and the maximum acceleration error reduced by \ensuremath{1.2\!\ \text{m/s}^2}. Compared to traditional controllers, the A-FTSMC controller offers faster convergence and less vibration, reducing the stabilization time by \ensuremath{27.3\%\!\ }.

\end{abstract}


\section{Introduction}

\IEEEPARstart{W}{ith} the rapid development of mobile communication and sensor technology, significant progress has been witnessed in autonomous driving \cite{yang,al}. However, fully autonomous driving still faces challenges, such as complex traffic environments, underdeveloped infrastructure, and legal restrictions \cite{loke}. As a transitional solution, human-machine co-driving systems have become mainstream. These systems enable collaboration between the driver and ADAS, providing functions like automatic vehicle-following, lane-keeping, and emergency avoidance, thereby enhancing driving safety and comfort \cite{ats,mus}. In this context, vehicle-following assistance has become a core functionality of ADAS, aimed at maintaining safe relative distances across various scenarios \cite{ben}. In these human-machine co-driving scenarios, both the driver and ADAS share control authority over the vehicle's acceleration. Due to the uncertainty of the driver's state, such as fatigue and decreased attention level, their actions may exhibit delays or sudden changes, and current assistance systems struggle to quickly adjust their vehicle-following strategies. Moreover, existing vehicle-following control algorithms often fail to maintain system stability when dealing with uncertainty and nonlinear parameter changes, leading to an increased risk of instability and collisions \cite{hu}. To address this, this work proposes an advanced human-machine co-driving system with robust adaptive control strategies that model driver state uncertainty, improving system stability in complex environments. 

Building a human-machine co-driving system starts with developing a vehicle-following model that monitors the driver's behavior in real time. This model serves as the basis for adjusting vehicle-following strategies when abnormal situations arise. Among the widely used models are the Optimal Velocity Model (OVM) and the Intelligent Driver Model (IDM) \cite{gho,aw}. OVM adjusts the vehicle’s speed according to the relative distance and speed of the preceding vehicle, helping to maintain smooth traffic flow during stable driving conditions \cite{fis,mn}. IDM incorporates more comprehensive factors, such as desired speed, acceleration, and safe following distance, allowing for precise control \cite{alex}. However, both OVM and IDM rely on fixed parameters, such as preset driver reaction time, which limits their adaptability in scenarios where the driver’s reaction time undergoes sudden changes \cite{shang}.

To address the limitations in robustness caused by the dependence on static parameters, this work focuses on integrating real-time driver state estimation into vehicle-following models. One common driver state estimation method collects historical data on driver actions to predict future behavior, enabling the system to anticipate driver response speed \cite{bo}. Alternatively, driver state estimation based on physiological signals, such as muscle electrical signals and facial expressions, offers a more direct method \cite{quyou,lie}. By utilizing neural networks to extract physiological features in real time \cite{si,ans}, this approach requires less data processing, allowing for faster adaptation to changes in driver attention or fatigue levels \cite{fuji,du}. In this work, intelligent driver model is enhanced by integrating real-time extraction of eyelid closure and mouth-opening features, allowing for a more precise evaluation of driver states within the vehicle-following system \cite{dongrui}.

To ensure that ADAS intervenes appropriately based on real-time driver state estimation, this work proposes a dynamic control allocation strategy to ensure smooth transitions of control authority. Current strategies can be categorized into several approaches. One approach is the game-theoretic strategy, resolving conflicts between the driver and the system during control transitions \cite{zhou,fang}. Another approach involves rule-based dynamic adjustments, relying on predefined rules and parameters to adjust control authority in real time \cite{yuan}. For instance, Zhang et al. designed an Allocation Domain method that uses simulated annealing algorithms to gradually increase ADAS control when the driver shows signs of fatigue or distraction \cite{ziyu}. Additionally, with advancements in sensor technology, real-time feedback mechanisms based on physiological signals have been integrated into allocation strategies. By monitoring physiological characteristics, systems can more accurately detect driver fatigue or distraction \cite{mar}. Steven et al. developed a feedback system that dynamically adjusts the control authority curve according to real-time physiological data \cite{steven,li}. Compared to traditional strategies that rely on predefined rules, the proposed strategy combines the flexibility of rule-based methods with the precision of physiological signal feedback. The system progressively adjusts ADAS intervention through a smooth, non-linear transition based on changes in the driver’s state, avoiding abrupt vehicle speed fluctuations caused by sudden control shifts \cite{fang}. 

To avoid instability caused by system parameter changes during control authority transition, this work requires an effective nonlinear control method \cite{peng}. As classic nonlinear control methods, Model Predictive Control (MPC) adjusts control strategies based on predictions of future system states, making it effective for managing complex constraints \cite{al}. \(H_\infty\) control focuses on enhancing robustness against disturbances. For example, Zhang et al. proposed a Lyapunov-Krasovskii-based \(H_\infty\) controller, which reduces vehicle-following errors by ensuring platoon convergence under various driving conditions \cite{bo}. However, \(H_\infty\) control presents challenges in time-varying systems, leading to overly conservative control actions and suboptimal performance \cite{ela}. Additionally, MPC struggles with delayed responses and reduced accuracy when faced with frequent external disturbances \cite{xu}.

Sliding Mode Control (SMC) has found extensive use in vehicle dynamics control because of its strong robustness against disturbances \cite{senel}. However, conventional SMC introduces undesirable chattering near the sliding surface, affecting the smoothness of acceleration, braking, and overall driving safety \cite{utk}. To address these issues, several improved versions of SMC have been proposed. Higher-order sliding mode control, which incorporates higher-order derivatives in the sliding surface design, effectively reduces chattering, making it suitable for high-speed and emergency response scenarios \cite{gian}. Terminal sliding mode control ensures that system states reach equilibrium within a finite time frame \cite{em,naj}. Adaptive Sliding Mode Control dynamically adjusts the sliding surface parameters in real time, allowing it to efficiently manage parameter uncertainties \cite{park}. By combining adaptive control with fast terminal sliding mode control, the system gains an improved capacity to handle nonlinearities and disturbances \cite{gao}. In this work, an integral sliding surface (ISM) is introduced alongside the non-singular terminal sliding surface, which dynamically adapts the control law by accounting for cumulative errors over time \cite{xin}. This two-layer mechanism leverages historical error data for more precise control adjustments, making it suitable for the fast and stable control required in human-machine co-driving systems. The main contributions of this paper are listed as follows:
\begin{enumerate}
    \item To address the inability of fixed-parameter vehicle-following models to simulate driver behavior under time-varying states, this study introduces driver reaction time as a dynamic adjustment factor based on the intelligent driver model. By monitoring facial features in real time, the system adjusts following distance and acceleration, providing a more precise simulation of driver behavior.
    \item To address the slow response and instability of existing vehicle-following control methods under driving authority switching, this study designs a two-layer adaptive non-singular terminal sliding mode controller by introducing an integral sliding mode surface. The proposed controller compensates for system uncertainties, ensuring less vibration and finite time stability.
    \item To optimize control authority distribution between the driver and ADAS, a dynamic control authority allocation strategy that considers real-time driver states is proposed. By incorporating fuzzy logic, control authority is smoothly adjusted according to a predefined control weight curve. The proposed strategy has been validated using real driver video data, demonstrating effective collaborative control under complex scenarios.
\end{enumerate}

The remaining sections are structured as follows: Section II introduces the modelling process of the vehicle following model and driver reaction time determination model. Section III presents the framework of HM-VFAS and the construction of the A-FTSMC. Section IV discusses the experimental setup and simulation results under different driver states and scenarios. Finally, Section V provides the conclusion of the paper and discusses potential directions for future research.

\section{Problem Formulation}
\begin{figure*}[t]
    \centering
    \includegraphics[width=6.7in]{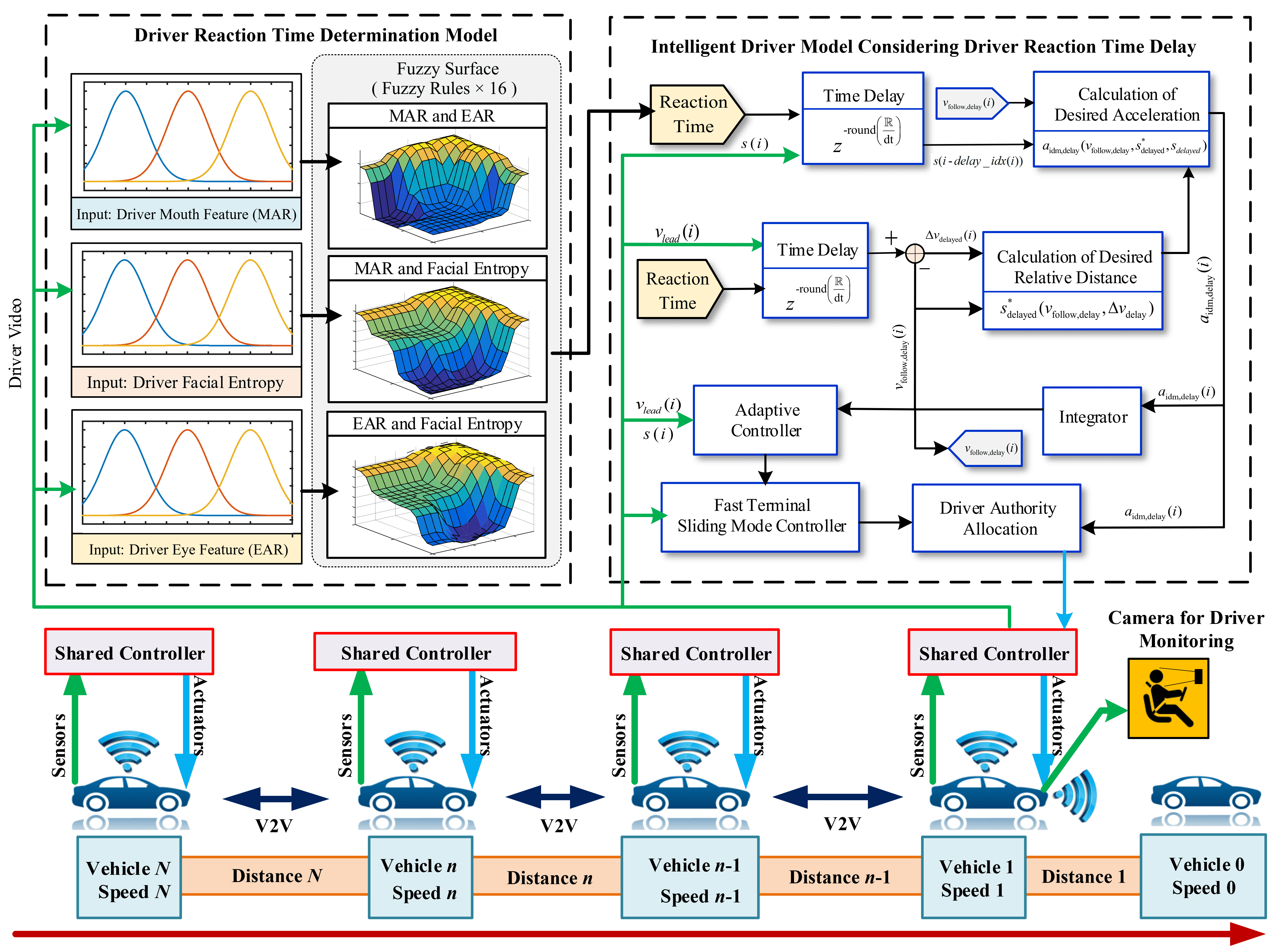}
    \caption{The structure of vehicle following behaviour model and driver reaction time determination model.}
    \label{fig:FC-IDM}
\end{figure*}
In human-machine co-driving vehicle-following scenarios, both the driver and the ADAS jointly control the vehicle’s acceleration to maintain safe distance. However, the driver’s state may fluctuate due to fatigue or distraction, manifesting as delayed or abrupt actions. Traditional vehicle-following assistance systems often lack the ability to adapt to such real-time changes in driver behavior.

To address these issues, it is essential to develop a more flexible vehicle-following assistance system that can incorporate driver state variations and adaptive control strategies in real-time. This work proposes two models in Fig. \ref{fig:FC-IDM}: (1) an intelligent driver model (IDM) considering driver reaction time delay to simulate the time-varying behavior of drivers under different states. (2) a fuzzy logic-based driver reaction time determination model that evaluates reaction time using real-time driver visual data.

\subsection{The IDM Considering Driver Reaction Time Delay}

The IDM is widely used in traffic flow and vehicle behavior simulations, describing driver acceleration and deceleration under varying conditions\cite{mn}. Its adaptability to different scenarios makes it an basic model for evaluating driver behavior in human-machine co-driving systems. The model operates by first calculating the desired distance \(s^*(i)\) based on the current speed \(v_{\text{follow}}(i)\), the time headway \(T\), and the relative speed \(\Delta v(i)\). This desired distance is then used to determine the desired acceleration \(a_{\text{idm}}(i)\). The core equations of the IDM model are as follows:

\begin{equation} 
\left\{
\begin{aligned}
&\Delta v(i) = v_{\text{follow}}(i) - v_{\text{lead}}(i) \\
&s(i) = d_{\text{follow}}(i) \\
&s^*(i) = s_0 + v_{\text{follow}}(i) \cdot T + \frac{v_{\text{follow}}(i) \cdot \Delta v(i)}{2 \sqrt{a_{\text{max}} \cdot b}} \\
&a_{\text{idm}}(i) = a_{\text{max}} \left( 1 - \left( \frac{v_{\text{follow}}(i)}{v_{\text{max}}} \right)^4 - \left( \frac{s^*(i)}{s(i)} \right)^2 \right) \\
&v_{\text{follow}}(i) = v_{\text{follow}}(i-1) + a_{\text{idm}}(i) \cdot \mathrm{d}t \\
&d_{\text{follow}}(i) = d_{\text{follow}}(i-1) + v_{\text{follow}}(i-1) \cdot \mathrm{d}t
\end{aligned}
\right.
\tag{1}
\end{equation}

where \(s(i)\) and \(d_{\text{follow}}(i)\) is the current relative distance between vehicles which can be obtained by V2V communication.  \(v_{\text{follow}}(i)\) and \(v_{\text{lead}}(i)\) are the speeds of the following and leading vehicles respectively, \(s_0\) is the minimum safety distance, \(T\) is the time headway, \(a_{\text{max}}\) is the maximum acceleration, \(b\) is the comfortable deceleration, and \(\text{d}t\) is the time step.

To simulate driver behavior accurately under time-varying driver states, the driver's reaction time delay is integrated into the IDM model, resulting in the intelligent driver model considering reaction time delay (IDM-RTD).  The process of integrating the reaction time delay into the IDM model involves several steps:

\begin{figure*}[t]
    \centering
    \includegraphics[width=\textwidth]{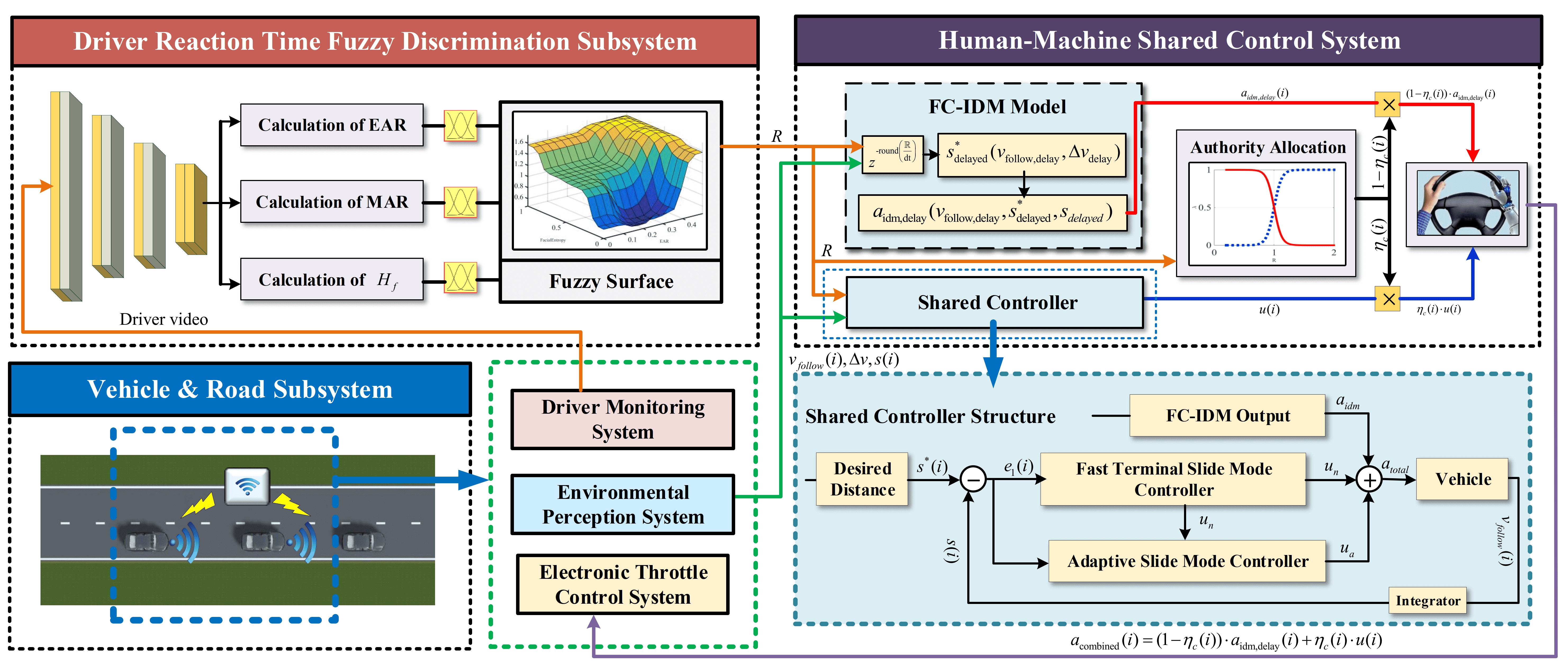}
    \caption{Control framework of human-machine shared vehicle-following assistance system.}
    \label{fig:ControlFramework}
\end{figure*}

First, calculate the delay index to account for the driver's reaction time delay. The delay index can be calculated by the following equations, where \( R \) represents the reaction time and \( \text{d}t \) is the time step:
\begin{equation}
\text{delay\_idx}(i) = \text{round} \left( \frac{R}{\text{d}t} \right) \setcounter{equation}{2}\label{eq2}
\end{equation}

Next, determine the delayed relative speed and distance. The relative speed \(\Delta v_{\text{delayed}}(i)\) and the vehicle distance \(s_{\text{delayed}}(i)\) at the delayed index are calculated using the following formulas:
\begin{align}
\Delta v_{\text{delayed}}(i) &= v_{\text{follow}}(i - \text{delay\_idx}(i)) \notag\\
&\quad- v_{\text{lead}}(i - \text{delay\_idx}(i)) \label{eq3} \\
s_{\text{delayed}}(i) &= d_{\text{follow}}(i - \text{delay\_idx}(i)) \label{eq4}
\end{align}

Then, compute the desired distance under delayed conditions. The desired distance \(s^*_{\text{delayed}}(i)\) is calculated as:
\begin{align}
s^*_{\text{delayed}}(i) &= s_0 + v_{\text{follow}}(i - \text{delay\_idx}(i)) \cdot T \notag \\
&\quad + \frac{v_{\text{follow}}(i - \text{delay\_idx}(i)) \cdot \Delta v_{\text{delayed}}(i)}{2 \sqrt{a_{\text{max}} \cdot b}} \label{eq5}
\end{align}

Finally, calculate the delayed acceleration. The delayed acceleration \(a_{\text{idm,delay}}(i)\) is calculated as:
\begin{align}
a_{\text{idm,delay}}(i) &= a_{\text{max}} \left( 1 - \left( \frac{v_{\text{follow}}(i - \text{delay\_idx}(i))}{v_{\text{max}}} \right)^4 \right. \notag \\
&\quad \left. - \left( \frac{s^*_{\text{delayed}}(i)}{s_{\text{delayed}}(i)} \right)^2 \right) \label{eq6}
\end{align}

\subsection{Driver Reaction Time Determination Model}\label{section2}

Driver reaction time can be estimated by physiological monitoring (heart rate, EEG), behavioral analysis (steering, pedal force) \cite{rah}, and visual feature extraction. Vision-based methods are popular in shared control systems for their non-invasive, real-time capabilities \cite{quyou}, using cameras to analyze facial features like eyelid closure, blink frequency, and head posture to assess fatigue and reaction time \cite{yansha}.

Our method applies the facial localization network from \cite{you} to extract driver facial coordinates, utilizing convolutional layers with batch normalization, leaky ReLU, and max-pooling. Dlib toolkit extracts facial feature points to compute vectors like the Eye Feature Vector (EFV), Mouth Feature Vector (MFV), and entropy \(H_F(X)\), capturing facial movement complexity. EFV is derived from eye feature distances, MFV from mouth movements, while \(H_F(X)\) reflects variability in facial motion. Details are provided in Appendix \ref{tiqu}.

The extracted Feature Vectors are fed into a fuzzy inference system, which applies fuzzy rules to infer the driver’s reaction time from the input features. This inferred reaction time is then used to adjust the human-machine shared control system, ensuring safe and efficient vehicle operation.

The driver's reaction time \(R\) is calculated using the following fuzzy inference process:
\begin{equation}
R = \frac{\sum_{j=1}^{n} w_j \cdot y_j}{\sum_{j=1}^{n} w_j}
\end{equation}
where \(w_j\) is the activation strength of the \(j\)-th fuzzy rule, and \(y_j\) is the output of the \(j\)-th rule. The membership functions for EFV, MFV, and \(H_F(X)\) are defined using Gaussian functions:

\begin{equation}
\mu(x) = \exp \left( -\frac{(x - c)^2}{2\sigma^2} \right)
\end{equation}

The fuzzy rule base consists of \ensuremath{16} fuzzy rules. The overall output reaction time \(R\) is determined using the centroid defuzzification method, providing a crisp value based on the weighted average of all rule outputs\cite{dongrui}.

\section{Human-Machine Shared Control System Design}
\subsection{Control Framework of Shared Vehicle Following System}

In this framework, the driver's facial characteristics are processed using the reaction time evaluation model outlined in Section II (B), enabling the identification of real-time reaction times. These times are then incorporated into a dynamic control allocation strategy, which adjusts the authority distribution between the driver and the control system, minimizing driver uncertainty and enhancing the interaction between human and machine. Building on this approach, the IDM-RTD described in Section II (A) operates alongside an adaptive sliding mode controller and a fast terminal sliding mode controller. These controllers together provide accurate acceleration commands to the vehicle-following system, ensuring both safety and stability.

Referring to the method in \cite{yin}, the human-machine shared control law is determined through the following equations:
\begin{equation}
a_{\text{combined}}(i) = (1 - \eta_c(i)) \cdot a_{\text{idm,delay}}(i) + \eta_c(i) \cdot h(i)
\end{equation}
where \( \eta_c \) is the authority allocation factor, defined as:
\begin{equation}
\eta_c =
\begin{cases}
0 & R < R_{\text{min}} \\
k_1 \left(1 + \tanh(k_2(R - R_{\text{mid}}))\right) & R_{\text{min}} \leq R \leq R_{\text{max}} \\
1 & R > R_{\text{max}}
\end{cases}
\end{equation}
where \( R \) represents the driver's reaction time, \( R_{\text{min}} \) and \( R_{\text{max}} \) are threshold values for the reaction time that determine the transition range, \( R_{\text{mid}} \) is the midpoint reaction time where the transition starts to change significantly, \( k_1 \) and \( k_2 \) are scaling parameters that control the shape of the transition function, \( a_{\text{idm,delay}}(i) \) is the delayed IDM acceleration, and \( h(i) \) is the control input from the adaptive controllers. The authority allocation factor \( \eta_c \) is dynamically adjusted based on the driver's reaction time. This design adapts the control system to the driver's state, improving safety and performance. When the driver's reaction time \( R \) is short, \( \eta_c \) is low, giving more control to the driver. As \( R \) increases, indicating fatigue, control shifts to the automated system to compensate for delayed reactions.

\subsection{Shared Controller Design}
To mitigate uncertainties caused by the switching of control authority between the driver and the ADAS system, an adaptive double-layer sliding mode controller is designed.

The state variables of the system are defined as:
\begin{align}
x_1(i) &= v_{\text{follow}}(i) \\
x_2(i) &= d_{\text{follow}}(i)
\end{align}

According the authority allocation strategy in \cite{yin} and \cite{fang}, the state space equation of the HM-VFAS system is constructed as:

\begin{align}
\begin{cases}
x_1(i) &= (1 - \eta_c(R)) \cdot a_{\text{max}} \left( 1 - \left( \frac{x_1(i - \text{delay\_idx})}{v_{\text{max}}} \right)^4 \right. \\
& \quad - \left( \frac{s^*_{\text{delayed}}(i)}{\epsilon_2(i - \text{delay\_idx})} \right)^2 \bigg) + \eta_c(R) \cdot h(i), \\
x_2(i) &= v_{\text{lead}}(i) - x_1(i)
\end{cases}
\end{align}
where \( \eta_c(R) \) is the authority allocation factor, \( a_{\text{max}} \) is the maximum acceleration, \( v_{\text{max}} \) is the maximum velocity, \( s^*_{\text{delayed}}(i) \) is the desired distance considering delay, and \( h(i) \) is the control input. The delay index \( \text{delay\_idx} \) accounts for the driver's reaction time delay.

To simplify the controller design process and ensure precise tracking performance, the tracking error signals are defined as:
\begin{align}
\epsilon_1(i) = s(i) - s^*(i)
\end{align}

These error signals represent the deviation of the actual distance from the desired distance. The system is then transformed into an error dynamics system to analyze the stability and design the controller. The error dynamics are given by:

\begin{align}
\begin{cases}
\epsilon_1(i) &= s(i) - s^*(i) \\
\epsilon_2(i) &= v_{\text{lead}}(i) - s^*(i) \\
& \quad - (1 - \eta_c(R)) \cdot a_{\text{max}} \left( 1 - \left( \frac{v_{\text{follow}}(i - \text{delay\_idx})}{v_{\text{max}}} \right)^4 \right. \\
& \quad \left. - \left( \frac{s^*_{\text{delayed}}(i)}{d_{\text{follow}}(i - \text{delay\_idx})} \right)^2 \right) - \eta_c(R) \cdot h(i)
\end{cases}
\label{eq15}
\end{align}
where \( \epsilon_1(i) \) and \( \epsilon_2(i) \) represent the errors in distance and velocity, respectively. The terms involving \( \eta_c(R) \) and \( h(i) \) illustrate the contribution of the control input to the system dynamics.

The error dynamics system can be simplified as equation (\ref{16})-(\ref{18}):
\begin{align}
\begin{cases}
\dot{\epsilon}_1(i) &= \epsilon_2(i) \\
\dot{\epsilon}_2(i) &= \chi(i) + \gamma(i) h(i)
\end{cases}
\label{16}
\end{align}
\begin{figure*}[!t]
\begin{align}
\chi(i) &= v_{\text{lead}}(i) - (1 - \eta_c(R)) \cdot a_{\text{max}} \left( 1 - \left( \frac{v_{\text{follow}}(i - \text{delay\_idx})}{v_{\text{max}}} \right)^4 - \left( \frac{s^*_{\text{delayed}}(i)}{d_{\text{follow}}(i - \text{delay\_idx})} \right)^2 - \ddot{s}^*(i) \right) \label{17}\\
\gamma(i) &= -\eta_c(R)\label{18}
\end{align}
\end{figure*}
\textbf{Assumption 1.} Consider the system with uncertainties:
\begin{equation}
|\chi(i)| \leq \bar{\chi} = \chi_0 + \chi_1 |\epsilon_1| + \chi_2 |\epsilon_2|
\end{equation}
The uncertainties \( \chi(i) \) and \( \gamma(i) \) are bounded as follows:
\begin{equation}
0 < K_m \leq \gamma(i) \leq K_M
\end{equation}
where \(\chi_0\), \(\chi_1\), \(\chi_2\), \(K_m\), and \(K_M\) are positive constants. It is assumed that \(K_m\) and \(K_M\) are known. Robust and adaptive controllers are designed for scenarios where \(\chi\) is either known or unknown. In the subsequent sections, \(\chi(x)\) and \(\gamma(x)\) are referred to simply as \(\chi\) and \(\gamma\).

\subsubsection{Fast Terminal Sliding Mode Controller Design}\par
In order to guarantee that the system reaches a steady state within a finite time, the initial and most crucial step is the design and construction of a fast terminal sliding mode controller. This controller ensures swift convergence to the desired state, optimizing system performance and stability.

\textbf{Theorem 1.} The designed controller can ensure that the error system (\ref{eq15}) achieves finite-time stability and the sliding surface does not exhibit convergence stagnation at the pole.

The designed controller \(h_n\) is as follows:
\begin{align}
h_n &= -\frac{1}{K_m} ( \alpha_2 \psi_1 + (B_1 + B_2 e^{-at}) \alpha_1 \text{sign}(\psi_1) \notag\\
& \quad+ \frac{1}{\beta q} |\epsilon_2|^{2-q} \text{sign}(\epsilon_2) )
\label{eq21}
\end{align}
where \(\alpha_1 = \bar{\chi} + \varsigma, \alpha_2 \geq 0, \varsigma\) is a small positive constant.

In the proposed control strategy, an Exponential Decline Switching Gain is introduced to mitigate the chattering effect:

\begin{equation}
A(t) \alpha_1 = (B_1 + B_2 e^{-at}) \alpha_1
\end{equation}
where \(\alpha_1 > \phi\), \(a > 0\), \(B_1 \geq 1\), and \(B_2 > 0\). This design enhances the system's convergence rate during the initial phase of reaching the sliding surface, while simultaneously reducing chattering amplitude without compromising the system's robustness.

\textbf{Lemma 1.}\label{lemma1} Consider the continuous autonomous system\cite{yu}:
\begin{equation}
\dot{x} = f(x), \quad f(0) = 0, \quad x \in \mathbb{R}^n
\end{equation}
where \( f: \mathbb{R}^n \rightarrow \mathbb{R}^n \) is locally Lipschitz.

Let \( V(x) \) be a continuous function defined in a neighborhood \( D \subseteq \mathbb{R}^n \) of the origin, and suppose it satisfies the following conditions:
1. \( V(x) \) is positive definite;
2. \( \dot{V}(x) + \rho V^\sigma(x) \leq 0, \quad \forall x \in D, \quad \rho > 0, \quad \sigma \in (0,1) \).

Then the origin of the system is locally finite-time stable. If \( D = \mathbb{R}^n \) and \( V(x) \rightarrow \infty \) as \( x \rightarrow \infty \), then the origin of the system is globally finite-time stable. The settling time \( T_s \) can be estimated as:
\begin{equation}
T_s \leq \frac{V(x_0)^{1-\sigma}}{\rho (1-\sigma)}
\label{eq24}
\end{equation}

Furthermore, if \( V(x) \) satisfies:
\begin{equation}
\dot{V}(x) + \theta V(x) + \rho V^\sigma(x) \leq 0, \quad \theta > 0
\end{equation}

The settling time \( T_s \) can be estimated as:
\begin{equation}
T_s \leq \frac{1}{\theta (1-\sigma)} \ln \left( \theta V(x_0)^{1-\sigma} + \frac{\rho}{\theta (1-\sigma)} \right)
\end{equation}

\textbf{Proof.} To achieve rapid stabilization, the first step is to construct the fast terminal sliding mode surface. The sliding surface \(\psi_1\) is defined as:
\begin{align}
\psi_1 = \epsilon_2 + \beta |\epsilon_2|^q \text{sign}(\epsilon_2)
\end{align}
where $ q = \delta + (1-\delta) \text{sign}(|\epsilon_2| - \epsilon) $ and $ 1 \leq \epsilon, 1 < \delta < 1.5 $.

The subsequent task involves formulating the Lyapunov function to evaluate the system's stability. This function is specified as follows:
\begin{align}
V = \frac{1}{2}\psi_1^2
\label{eq28}
\end{align}

Differentiating Lyapunov function (\ref{eq28}) with respect to time:
\begin{equation}
\dot{V} = \psi_1\dot{\psi_1}
\end{equation}

Substituting \( \dot{\psi_1} \) with the error terms:
\begin{equation}
\dot{\psi_1} = \epsilon_1 + \beta q |\epsilon_2|^{q-1} \epsilon_2 \text{sign}(\epsilon_2)
\end{equation}
\begin{equation}
\dot{\psi_1} = \epsilon_2 + \beta q |\epsilon_2|^{q-1} (\chi(i) + \gamma(i) h) \text{sign}(\epsilon_2)
\end{equation}

Further substituting controller (\ref{eq21}) and simplifying:
\begin{align}
\dot{V_1}& =\psi_1\left(\epsilon_{2}+\beta q|\epsilon_{2}|^{q-1}(\chi(i)+\gamma(i)h)\mathrm{sign}(\epsilon_{2})\right) \\
&=\psi_1\left(\epsilon_{2}+\beta q|\epsilon_{2}|^{q-1}\left(\chi(i)-\gamma(i)\frac{1}{K_{m}} (\alpha_{2}\psi_1\right.\right. \notag\\
&+(\nu_1+\nu_2e^{-at})\alpha_1\mathrm{sign}(\psi_1) \notag\\
&+\frac{1}{\beta q}|\epsilon_2|^{2-q}\text{sign}(\epsilon_2)\bigg)\bigg)\text{sign}(\epsilon_2)\notag
\end{align}

Considering the inequalities \(0 < K_m \leq \gamma(i) \leq K_M\):
\begin{align}
\dot{V_1} &= -\beta q |\epsilon_2|^{q-1} \frac{\gamma}{K_m} \alpha_2 \psi_1^2 + \psi_1 \epsilon_2 \notag \\
&\quad - \frac{\gamma}{K_m} |\psi_1| |\epsilon_2| \notag \\
&\quad + \beta q |\epsilon_2|^{q-1} \left( \chi(i) - \frac{\gamma}{K_m} A(t) \alpha_1 |\psi_1| \right)
\end{align}

Considering the inequalities  \(A(t)\alpha_1\geq \bar{\chi} + \varsigma\):
\begin{align}
\dot{V} &\leq \psi_1\epsilon_2 - \beta q |\epsilon_2|^{q-1} \left( \alpha_2 \psi_1 + \varsigma \text{sign}(\psi_1) \right) \text{sign}(\epsilon_2)  \\
\dot{V} &\leq -\beta q \alpha_2 |\epsilon_2|^{q-1} \psi_1^2 - \beta q \varsigma |\epsilon_2|^{q-1} |\psi_1| 
\end{align}

Further simplification of the Lyapunov function's derivative yields:
\begin{align}
\dot{V} &\leq -\beta q \alpha_2 |\epsilon_2|^{q-1} V - \beta q \varsigma |\epsilon_2|^{q-1} \sqrt{2V}
\end{align}

The inequality from Lemma 1  can be applied as follows:
\begin{align}
\dot{V} &\leq -\beta q \alpha_2 |\epsilon_2|^{q-1} V^{1/2}
\end{align}

Based on the finite-time stability condition (\ref{eq24}), the following can be concluded:
\begin{align}
T &\leq \frac{V(0)^{1-\beta}}{\kappa(1-\beta)}
\end{align}

Therefore, the settling time \(T\) is given by:
\begin{align}
T &\leq \left( \frac{1}{2} \psi_1(0)^2 \right)^{1-1/2} \frac{\psi_1(0)}{2\kappa} = \frac{\psi_1(0)}{-2 \beta q \alpha_2 |\epsilon_2|^{q-1}}
\end{align}

Then, given the sliding surface's non-stagnation proof at the pole \( (\epsilon_1 \neq 0, \epsilon_1 = 0)\).

Substitute the controller (\ref{eq21}) into the system (\ref{eq15}) and take \( \epsilon_2 = 0 \):
\begin{align}
\dot{\epsilon}_{2}&=\chi-\frac{\gamma}{K_{m}}\left(\alpha_{2}\psi_1+A(t)\alpha_{1}\mathrm{sign}(\psi_1)\right.\notag\\&+\frac{1}{\beta q}|\epsilon_{2}|^{2-q}\mathrm{sign}(\epsilon_{2})
\end{align}

When $\epsilon_1 < 0$, \(\dot{\epsilon}_{2}\) can satisfy this inequality:
\begin{align}
\dot{\epsilon}_{2}& =-\alpha_{2}\frac{\gamma}{K_{m}}\psi_1+\chi-A\left(t\right)\alpha_{1}\frac{\gamma}{K_{m}} \notag\\
&=-\alpha_{2}\frac{\gamma}{K_{m}}\psi_1+\chi-\frac{\gamma}{K_{m}}\left(B_1+B_2e^{-at}\right)\left(\overline{\chi}+\varsigma\right) \\
&\leq\underbrace{-\alpha_{2}\psi_1}_{<0}+\underbrace{\chi-\overline{\chi}-\varsigma}_{<0}.\notag
\end{align}

When $\epsilon_1 > 0$, \(\dot{\epsilon}_{2}\) can satisfy this inequality:
\begin{align}
\dot{\epsilon}_{2}& =-\alpha_{2}\frac{\gamma}{K_{m}}\psi_1+\chi+A\left(t\right)\alpha_{1}\frac{\gamma}{K_{m}} \notag\\
&=-\alpha_{2}\frac{\gamma}{K_{m}}\psi_1+\chi+\frac{\gamma}{K_{m}}\left(B_1+B_2e^{-at}\right)\left(\overline{\chi}+\varsigma\right) \\
&\geq\underbrace{-\alpha_{2}\psi_1}_{>0}+\underbrace{\chi+\overline{\chi}+\varsigma}_{>0}.\notag
\end{align}

It can be inferred that the points where \(\{x_1 \neq 0, x_2 = 0\}\) are not attractors. Therefore, the movement along the sliding surface will not remain fixed at these points but will instead converge to zero within a finite time, ensuring that the system continues to evolve towards stability without stagnation.

The Proof is completed.\hfill $\square$
\newline
\subsubsection{Adaptive Sliding Mode Controller Design}\par
To address the uncertainties and disturbances in the system, a double-layer adaptive sliding mode control scheme is designed. This control strategy combines the advantages of integral sliding mode (ISM) and FTSMC. As the first layer, ISM compensates for the effects of system uncertainties and disturbances from the outset. As the second layer, FTSMC speeds up the convergence of system states towards the equilibrium point, providing rapid stabilization. This integrated approach estimates and compensates for unknown dynamics and disturbances, ensuring robust performance and improved system stability. Therefore, The shared controller \(h\) consists of an adaptive controller \(h_a\) together with the previously designed fast terminal sliding mode controller \(h_n\).

\begin{equation}
h = h_a + h_n
\end{equation}

The adaptive sliding surface \(\psi_1\)is constructed as follows:

\begin{equation}
\psi_1 = \epsilon_2 + z - e^{-\theta t} \left( \epsilon_2(0) + z(0) \right)
\end{equation}
\begin{equation}
\dot{z} = -h_n 
\end{equation}

The derivative of \( \psi_1 \) is given by:

\begin{align}
\dot{\psi}_1 &= \dot{\epsilon}_2 + \dot{z} + \Gamma = \chi + \gamma h - h_n + \Gamma\\
&\quad= \chi + \Gamma + (\gamma - 1) h_n + \gamma h_a\notag
\end{align}
\begin{equation}
\Gamma = \theta e^{-\theta t} \left( \epsilon_2(0) + z(0) \right)
\end{equation}

The adaptive sliding mode control law is then designed in Theorem 2 to ensure that the sliding surface \( \psi_1 \) converges to zero in finite time, compensating for the uncertainties and disturbances in the system.

\textbf{Assumption 2.} \label{lemma2} Assume \( \chi + (\gamma - 1) h_n \) is unknown and satisfies the following inequality:
\begin{equation}
\left| \chi + (\gamma - 1) h_n \right| \leq \xi_0 + \xi_1 |\epsilon_1| + \xi_2 |\epsilon_2|
\end{equation}
where \( \xi_0, \xi_1, \) and \( \xi_2 \) are unknown positive constants.

\textbf{Theorem 2.} \label{a2}For system (\ref{eq15}), the sliding surface \( \psi_1 \) converges to zero asymptotically with the following adaptive controller:
\begin{align}
h_a &= -\frac{1}{K_m} \left( k_3 \psi_1 + k_4 |\psi_1|^{p_2} \text{sign}(\psi_1) + |\Gamma| \text{sign}(\psi_1) \right. \notag \\
& \quad \left. + \left( \hat{\xi}_0 + \hat{\xi}_1 |\epsilon_1| + \hat{\xi}_2 |\epsilon_2| \right) \text{sign}(\psi_1) \right)
\label{eq49}
\end{align}

In equation(\ref{eq49})
\begin{align}
\dot{\hat{\xi}}_0 &= k_0 |\psi_1| \label{eq50}\\
\dot{\hat{\xi}}_1 &= k_1 |\psi_1| |\epsilon_1| \label{eq51}\\
\dot{\hat{\xi}}_2 &= k_2 |\psi_1| |\epsilon_2|
\label{eq52}
\end{align}
where \( \{k_0 \sim k_4\} \geq 0 \), \( 1 \geq p_2 > 0 \), and \( \hat{\xi}_i(0) > 0 \) (i = 0, 1, 2).

\textbf{Proof.} Assume \( \tilde{\xi}_i = \xi_i - \hat{\xi}_i \). Consider the following new Lyapunov function candidate:
\begin{equation}
V_2 = \frac{1}{2} \psi_1^2 + \frac{1}{2k_0} \tilde{\xi}_0^2 + \frac{1}{2k_1} \tilde{\xi}_1^2 + \frac{1}{2k_2} \tilde{\xi}_2^2
\end{equation}

The time derivative of \( V_2 \) is given by:
\begin{align}
\dot{V}_2 &= \psi_1 (\chi + \Gamma + (\gamma - 1)h_n + \gamma h_a)  \notag\\
&\quad-\left( \frac{1}{k_0} \tilde{\xi}_0 \dot{\tilde{\xi}}_0 \right) - \left( \frac{1}{k_1} \tilde{\xi}_1 \dot{\tilde{\xi}}_1 \right) - \left( \frac{1}{k_2} \tilde{\xi}_2 \dot{\tilde{\xi}}_2 \right)
\end{align}

Based on adaptive law (\ref{eq50})-(\ref{eq52}) and \( \hat{\xi}_i(0) > 0 \) :
\begin{align}
\dot{V}_2 &= \psi_1 (\chi + (\gamma - 1)h_n + \gamma \psi_1 h_a - |\psi_1| (\tilde{\xi}_0 + \tilde{\xi}_1 |\epsilon_1| + \tilde{\xi}_2 |\epsilon_2|)) \notag \\
&\leq |\psi_1| |\chi| + (\gamma - 1)h_n + \gamma \psi_1 h_a - |\psi_1| (\tilde{\xi}_0 + \tilde{\xi}_1 |\epsilon_1| + \tilde{\xi}_2 |\epsilon_2|) \notag \\
&\leq |\psi_1| (\xi_0 + \xi_1 |\epsilon_1| + \xi_2 |\epsilon_2|) + |\psi_1| |h_a| \notag\\
&\quad- \frac{\gamma}{K_m} \left( k_3 \psi_1^2 + k_4 |\psi_1|^{p_2 + 1} + |\Gamma| + \sum_{i=0}^2 \tilde{\xi}_i |\epsilon_i| \right) \notag \\
&= - \frac{\gamma}{K_m} \left( k_3 \psi_1^2 + k_4 |\psi_1|^{p_2 + 1} \right)
\label{eq52s}
\end{align}

Therefore, the equation (\ref{eq52s}) can be summarised as:
\begin{equation}
\dot{V}_2 \leq -\frac{\gamma}{K_m} \left( k_3 \psi_1^2 + k_4 |\psi_1|^{p_2 + 1} \right)
\end{equation}

According to Lemma 1, this ensures the asymptotic stability of the sliding surface \( \psi_1 \).

The Proof is completed.\hfill $\square$
\newline

The boundary layer method is frequently employed to mitigate chattering in sliding mode control systems. Instead of forcing the system to remain exactly on the sliding surface, this technique permits the system to function within a small neighborhood around it. Building upon this principle, a refined adaptive law was formulated to prevent the overestimation of adaptive gain, thereby ensuring that it converges gradually to the desired value over time.

When the boundary layer approach is applied to the adaptive controller, as indicated by equation (\ref{eq57}), it allows for a further reduction in the chattering amplitude. This occurs as the adaptive gains $\hat{\xi}_0, \hat{\xi}_1, \hat{\xi}_2$ are fine-tuned to their optimal values. As a result, the more refined A-FTSMC controller was constructed as follows:

\begin{align}\label{eq57}
h&=\frac1{K_m}\left(k_3\psi_1+k_4|\psi_1|^{p_2}\text{sat}(\psi_1)\right. \notag\\
&+|\Gamma|\text{sat}(\psi_1)+(\hat{\xi}_0+\hat{\xi}_1|\epsilon_1|+\hat{\xi}_2|\epsilon_2|)\text{sat}(\psi_1) \\
&-\frac{1}{K_m}(\alpha_2\psi_1+(B_1+B_2e^{-at})\alpha_1\text{sat}(\psi_1) \notag\\
&+\frac1{\beta q}|\epsilon_2|^{2-q}\text{sat}(\epsilon_2))\notag
\end{align}
where \(\text{sat}(\psi_1)\) is the saturation function defined as:
\begin{equation}
\text{sat}(\psi_1) = \begin{cases}
\frac{\psi_1}{\phi}, & \text{if } |\psi_1| \leq \phi \\
\text{sign}(\psi_1), & \text{if } |\psi_1| > \phi
\end{cases}
\end{equation}

The adaptive laws for the gains are adjusted to:
\begin{equation}
\dot{\hat{\xi}}_0 =
\begin{cases}
k_0 |\psi_1| \text{sign}(|\psi_1| - \phi), & \hat{\xi}_0 > \alpha_{\xi0} \\
\bar{k}_0, & \hat{\xi}_0 \leq \alpha_{\xi0},
\end{cases}
\end{equation}

\begin{equation}
\dot{\hat{\xi}}_1 =
\begin{cases}
k_1 |\psi_1| |\epsilon_1| \text{sign}(|\psi_1| - \phi), & \hat{\xi}_1 > \alpha_{\xi1} \\
\bar{k}_1, & \hat{\xi}_1 \leq \alpha_{\xi1},
\end{cases}
\end{equation}

\begin{equation}
\dot{\hat{\xi}}_2 =
\begin{cases}
k_2 |\psi_1| |\epsilon_2| \text{sign}(|\psi_1| - \phi), & \hat{\xi}_2 > \alpha_{\xi2} \\
\bar{k}_2, & \hat{\xi}_2 \leq \alpha_{\xi2},
\end{cases}
\end{equation}
where the positive constant \(\phi\) denotes the boundary layer, \(\{\bar{k}_0 \sim \bar{k}_2\}\) and \(\{\alpha_{\xi0} \sim \alpha_{\xi2}\}\) are positive constants.

\section{Simulation Considering Driver State}
\subsection{Experimental Scenario}
\begin{figure}[h]
    \centering
    \includegraphics[width=3.5in]{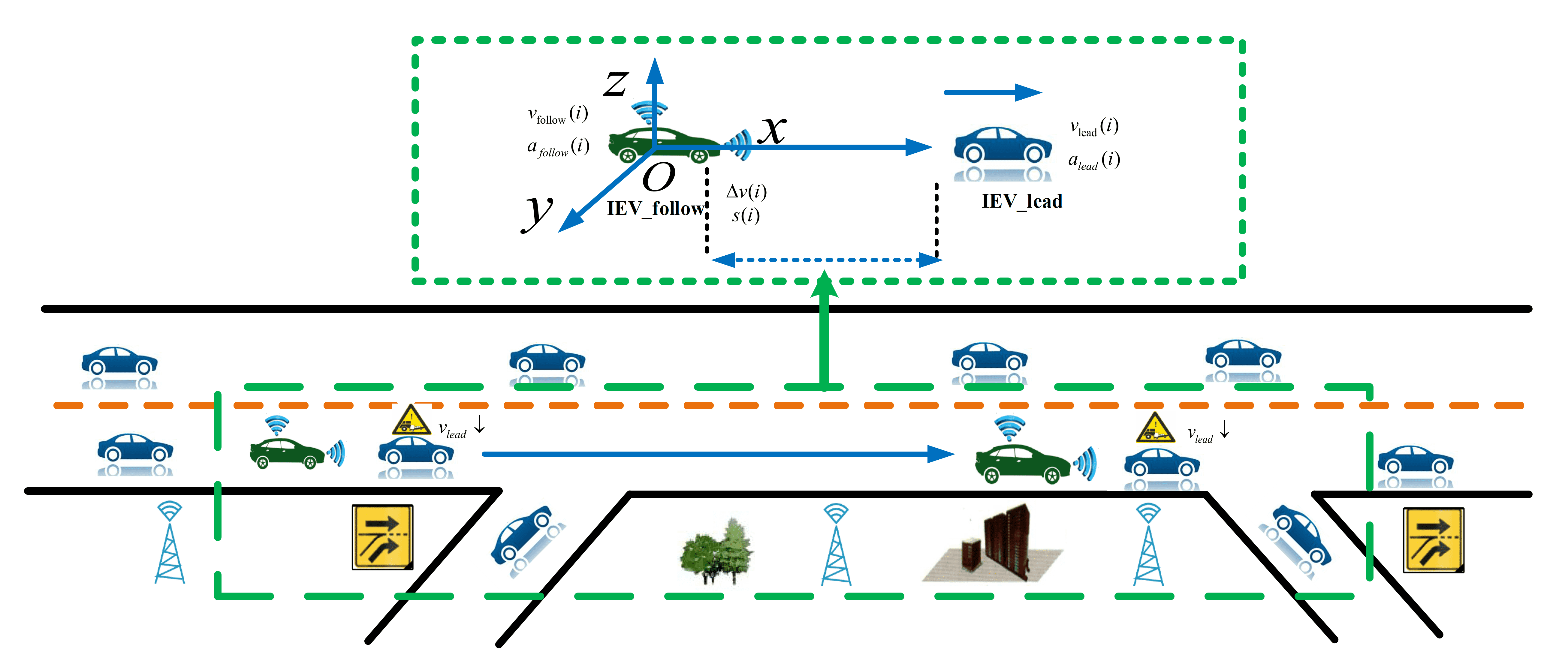}
    \caption{A scenario of ramp weaving area.}
    \label{fig:scenario}
\end{figure}

The capability of HM-VFAS system is verified in this section under a typical ramp weaving scenario (as shown in Fig. \ref{fig:scenario}). This scenario is chosen because vehicles entering and exiting the ramp cause the lead vehicle to frequently accelerate and decelerate. In such conditions, changes in the driver's reaction time significantly impact the vehicle-following performance.  

\begin{figure}[!b]
    \centering
    \includegraphics[width=3.5in]{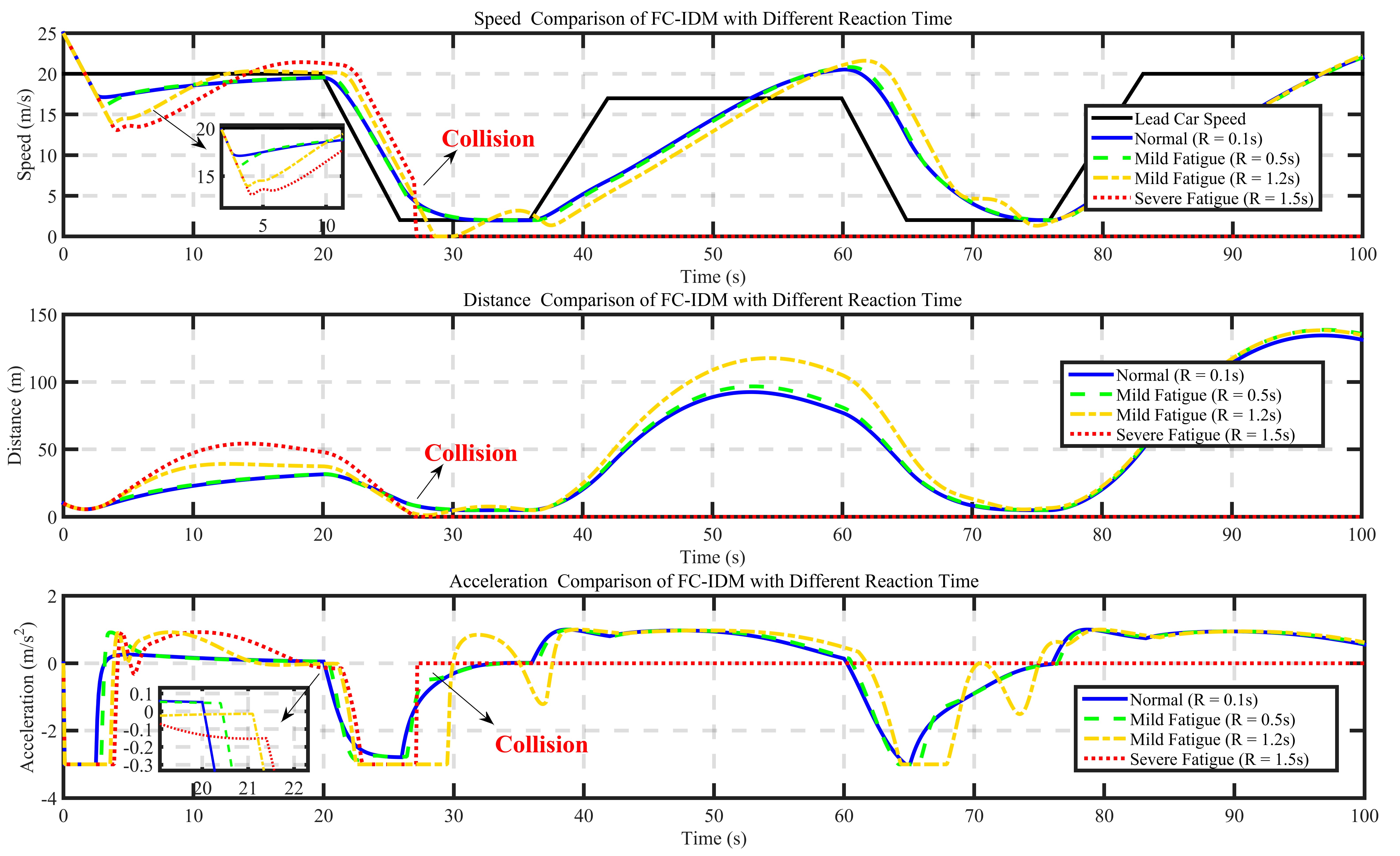}
    \caption{The output of the IDM-RTD model for
different driver states during acceleration and deceleration of the lead vehicle.}
    \label{fig:IDM}
\end{figure}

In this setup, the leading vehicle undergoes two distinct deceleration phases, followed by acceleration phases, which test the responsiveness and adaptability of the vehicle’s vehicle-following control system. The motion of the leading vehicle is described by the following set of equation (\ref{eq62}):
\begin{equation}
\begin{aligned}
v_{\text{lead}}(i) &=
\begin{cases}
20, & \text{if } t(i) = 0 \\
v_{\text{lead}}(i-1) - 3 \cdot \Delta t, & \text{if } t(i) \in [20, 26] \\
2, & \text{if } t(i) \in [26, 36] \\
v_{\text{lead}}(i-1) + 2.5 \cdot \Delta t, & \text{if } t(i) \in [36, 42] \\
v_{\text{lead}}(i-1) - 3 \cdot \Delta t, & \text{if } t(i) \in [60, 66] \\
2, & \text{if } t(i) \in [66, 76] \\
v_{\text{lead}}(i-1) + 2.5 \cdot \Delta t, & \text{if } t(i) \in [76, 82] \\
v_{\text{lead}}(i-1) + 2.5 \cdot \Delta t, & \text{if } t(i) \in [82, 92] \\
v_{\text{lead}}(i-1), & \text{if } t(i) \in [92, 100] \\
v_{\text{lead}}(i-1), & \text{otherwise}
\end{cases}
\end{aligned}
\label{eq62}
\tag{62}
\end{equation}

\subsection{Participants Description}
In this experiment, we recruited \ensuremath{21} volunteers aged between \ensuremath{20} and \ensuremath{45} with rich driving experience, including \ensuremath{6} females and \ensuremath{15} males. The participants were recorded in a controlled driving simulator environment while performing various driving tasks. These tasks simulated real-world driving conditions that could induce fatigue, such as long periods of monotonous driving and tasks with gradually increasing complexity. Real-time facial feature analysis and reaction time assessments were used to monitor and evaluate the participants' states. Throughout the experiment, we collected approximately \ensuremath{200} video segments and \ensuremath{3000} images. From this data, a representative video dataset was built, including different fatigue levels subset (such as sober, fatigued, and severely fatigued) and key scenarios segments with sudden changes in fatigue state. This subset was used to validate the functionality and adaptability of the HM-VFAS. All participants provided informed consent, and measures were taken to protect the privacy of the collected data.

\subsection{Simulation Under Different Driver Reaction Time}

Using a representative video subset from participants, simulations were conducted to study vehicle-following behavior under various driver states. These data were input into the IDM-RTD model to analyze how different reaction times influence speed, distance, and acceleration control.

Fig. \ref{fig:IDM} shows the IDM-RTD model output for various driver states. The speed comparison indicates that with a reaction time of \ensuremath{0.1\!\ \text{s}}, the following vehicle closely matches the lead vehicle’s speed. As reaction time increases to \ensuremath{0.5\!\ \text{s}}, \ensuremath{1.2\!\ \text{s}}, and \ensuremath{1.5\!\ \text{s}}, speed adjustments are delayed, especially at \ensuremath{1.5\!\ \text{s}}. The distance comparison shows a safe following distance of about \ensuremath{10\!\ \text{m}} at \ensuremath{0.1\!\ \text{s}}, but deteriorates as reaction time increases, with a collision at \ensuremath{1.5\!\ \text{s}}. The acceleration comparison reveals smooth acceleration with a maximum of \ensuremath{2.5\!\ \text{m/s}^2} at \ensuremath{0.1\!\ \text{s}}, but response becomes more erratic with longer reaction times. These results highlight the need for advanced controllers to compensate for delayed reactions, ensuring safe and stable vehicle-following behavior under varying driver conditions.

\begin{figure*}[!htbp]
    \centering
    \includegraphics[width=6.8in]{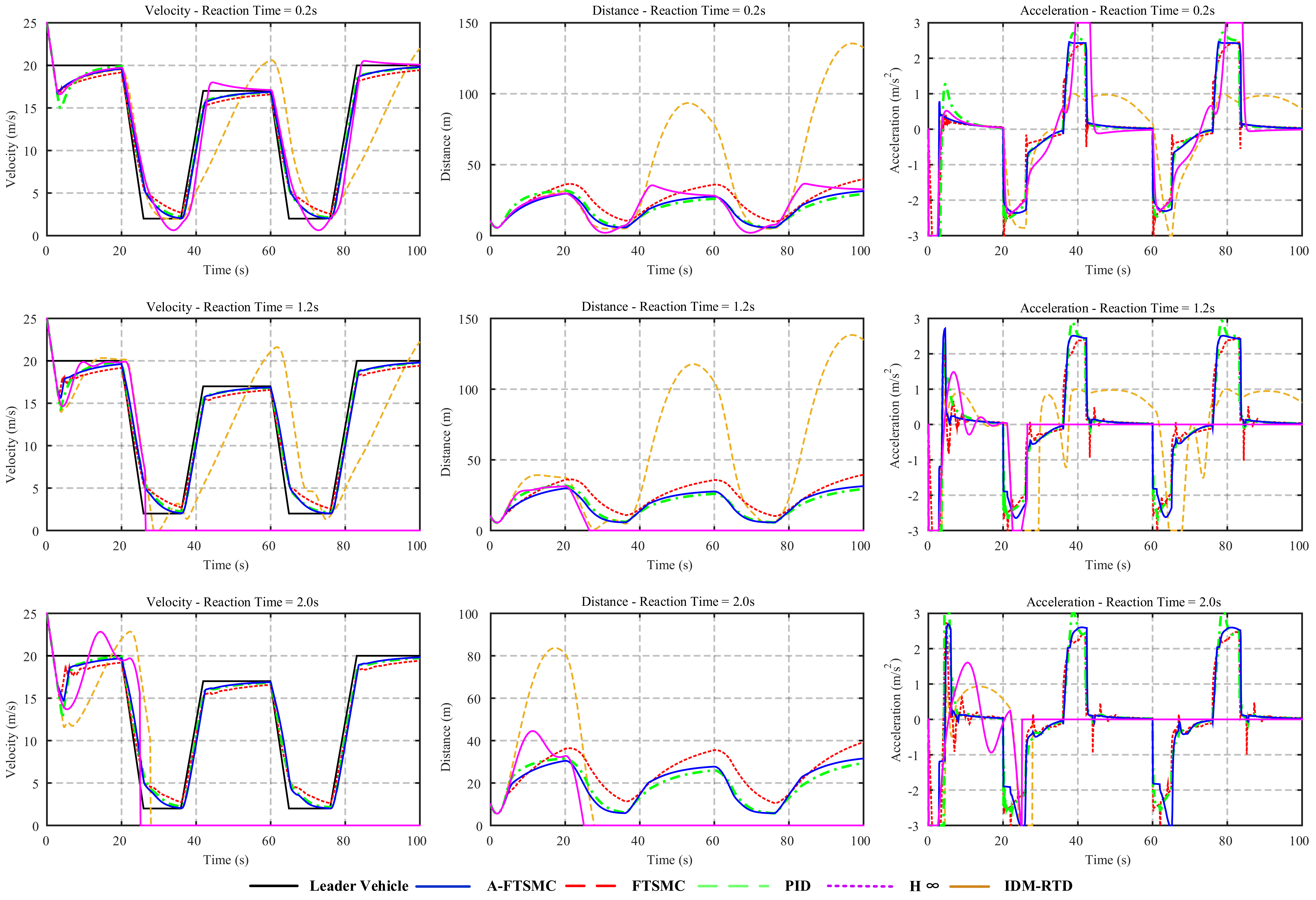}
    \caption{Speed, relative distance and acceleration of HM-VFAS under various driving conditions using different controller.}
    \label{fig:jgg}
\end{figure*}
\begin{figure}[!htbp]
    \centering
    \includegraphics[width=3.2in]{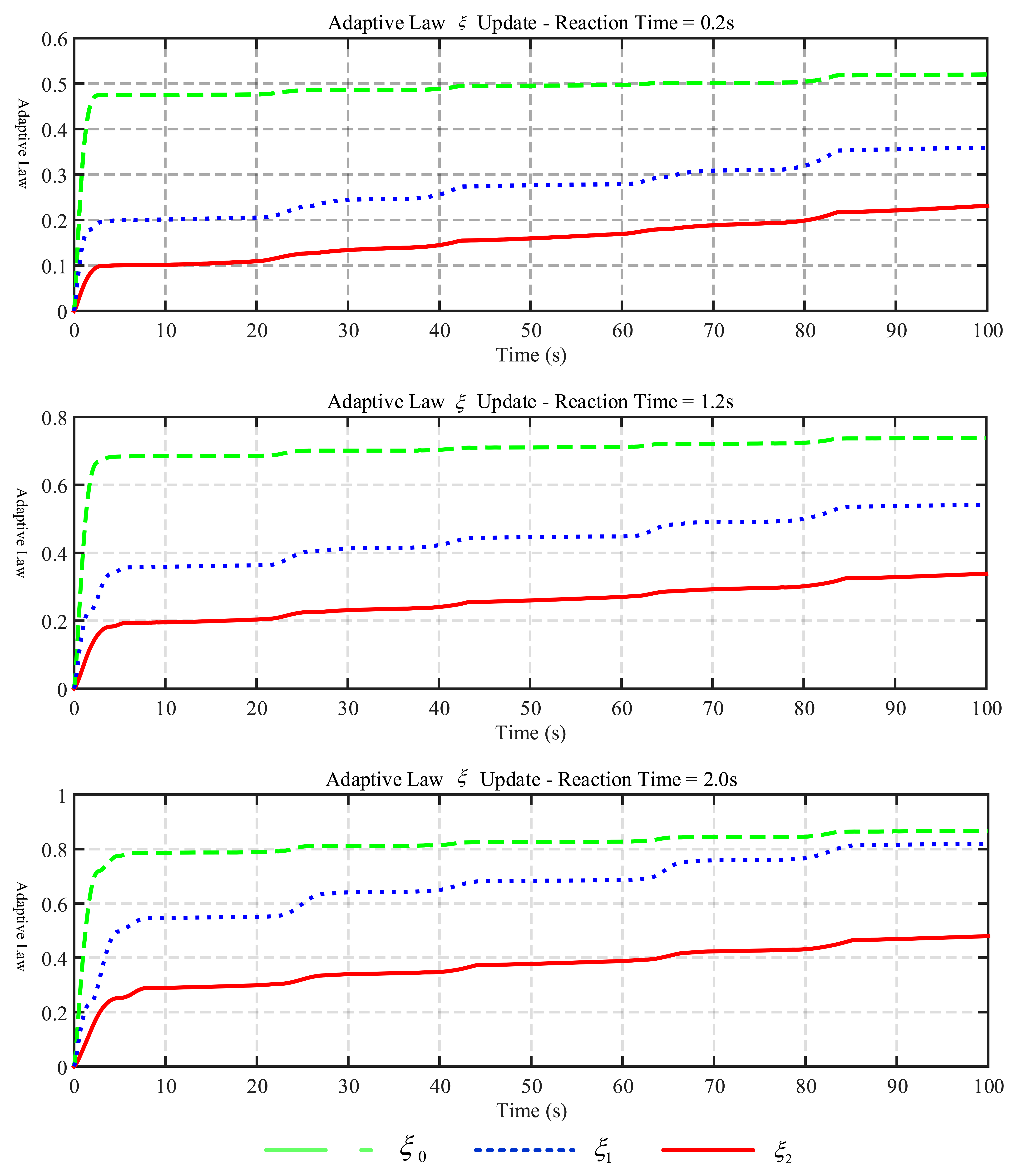}
    \caption{Adaptive law update of the A-FTSMC under different driver state.}
    \label{fig:ada1}
\end{figure}
\begin{figure}[!htbp]
    \centering
    \includegraphics[width=3.5in]{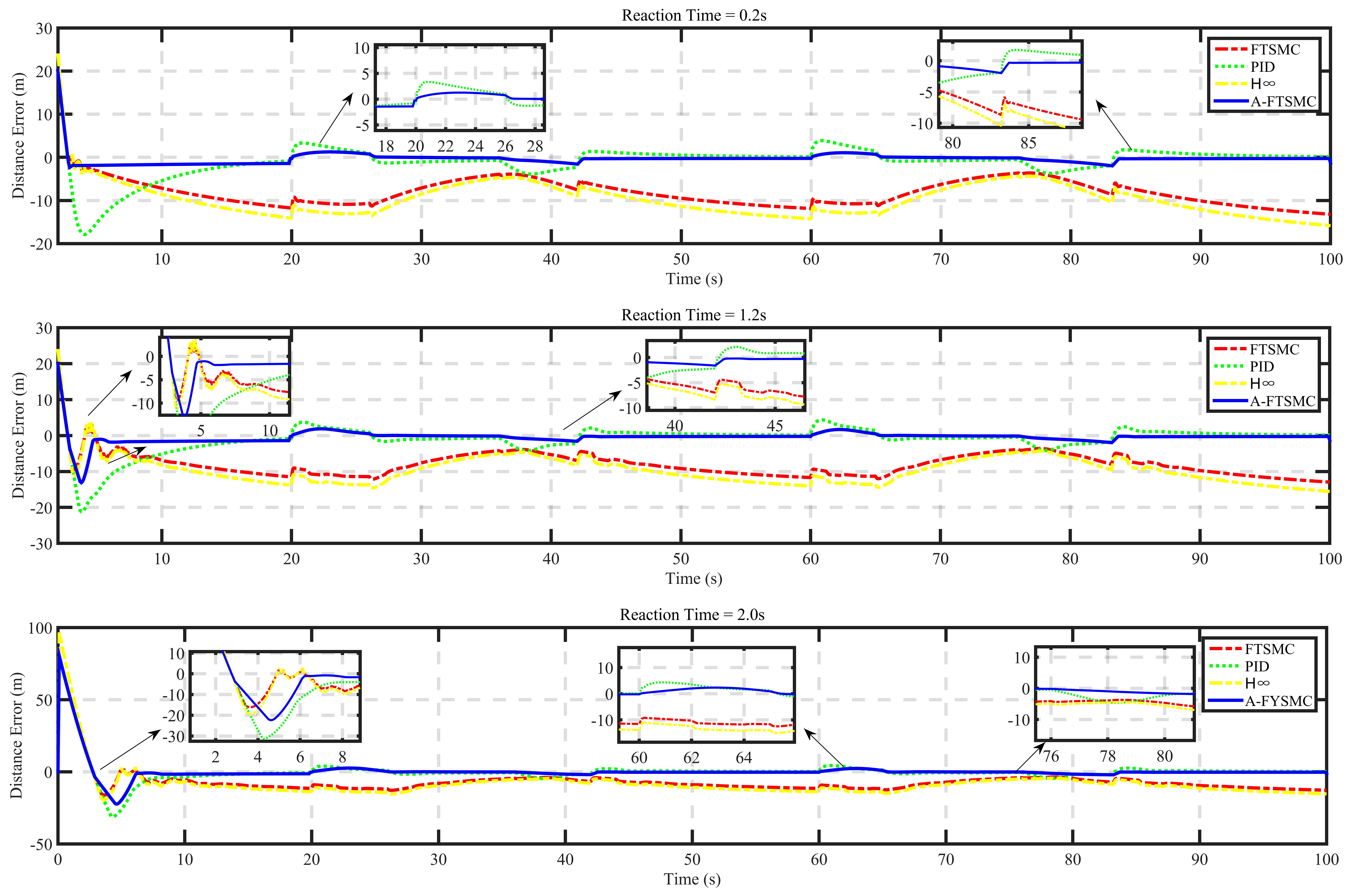}
    \caption{Error comparison of different controller at various driver reaction time.}
    \label{fig:err}
\end{figure}
\begin{figure*}[!htbp]
    \centering
    \includegraphics[width=6.5in]{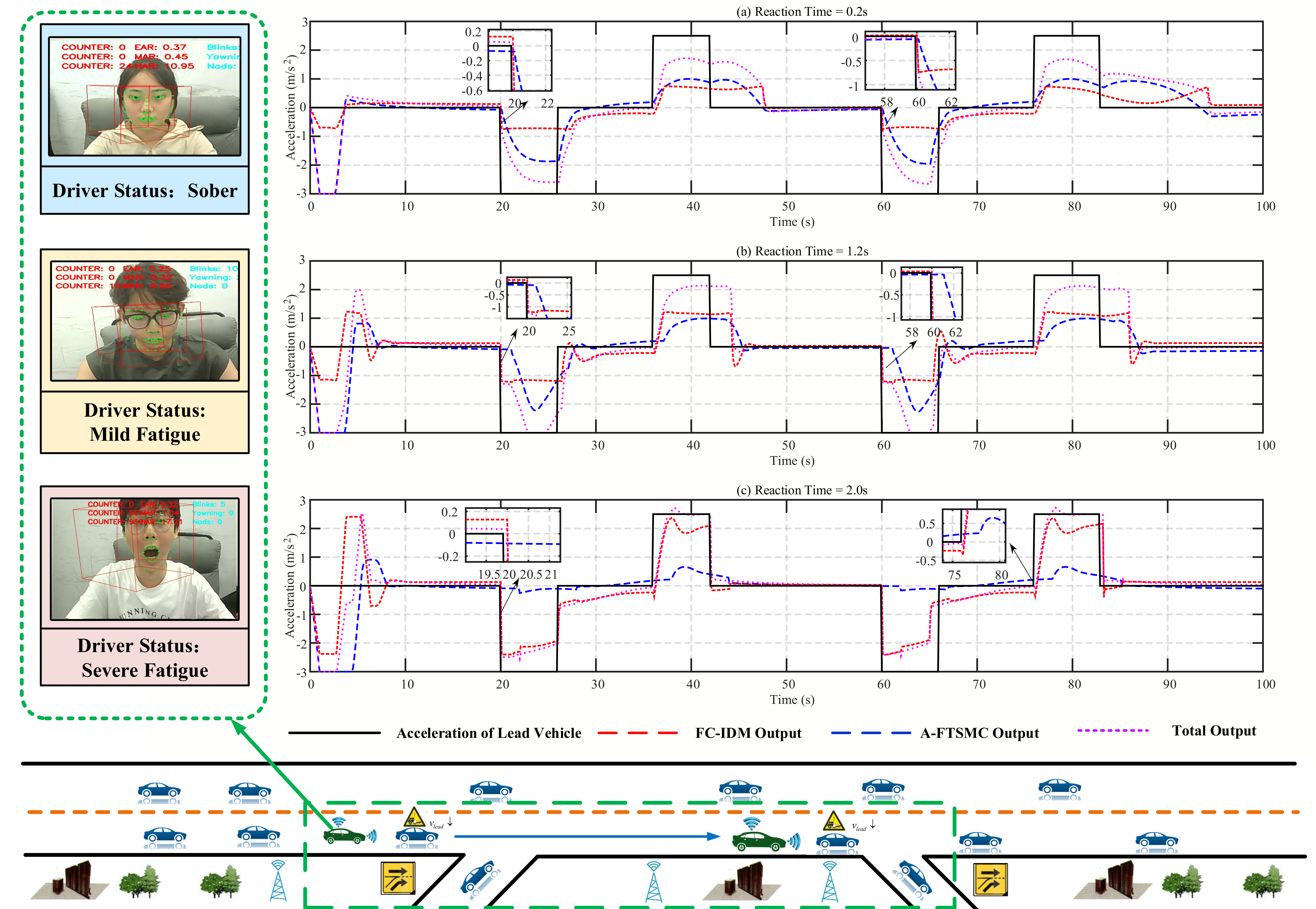}
    \caption{Acceleration of driver/assisted driving system output in different driver states.}
    \label{fig:acc}
\end{figure*}
\begin{figure}[!htbp]
    \centering
    \includegraphics[width=3.2in]{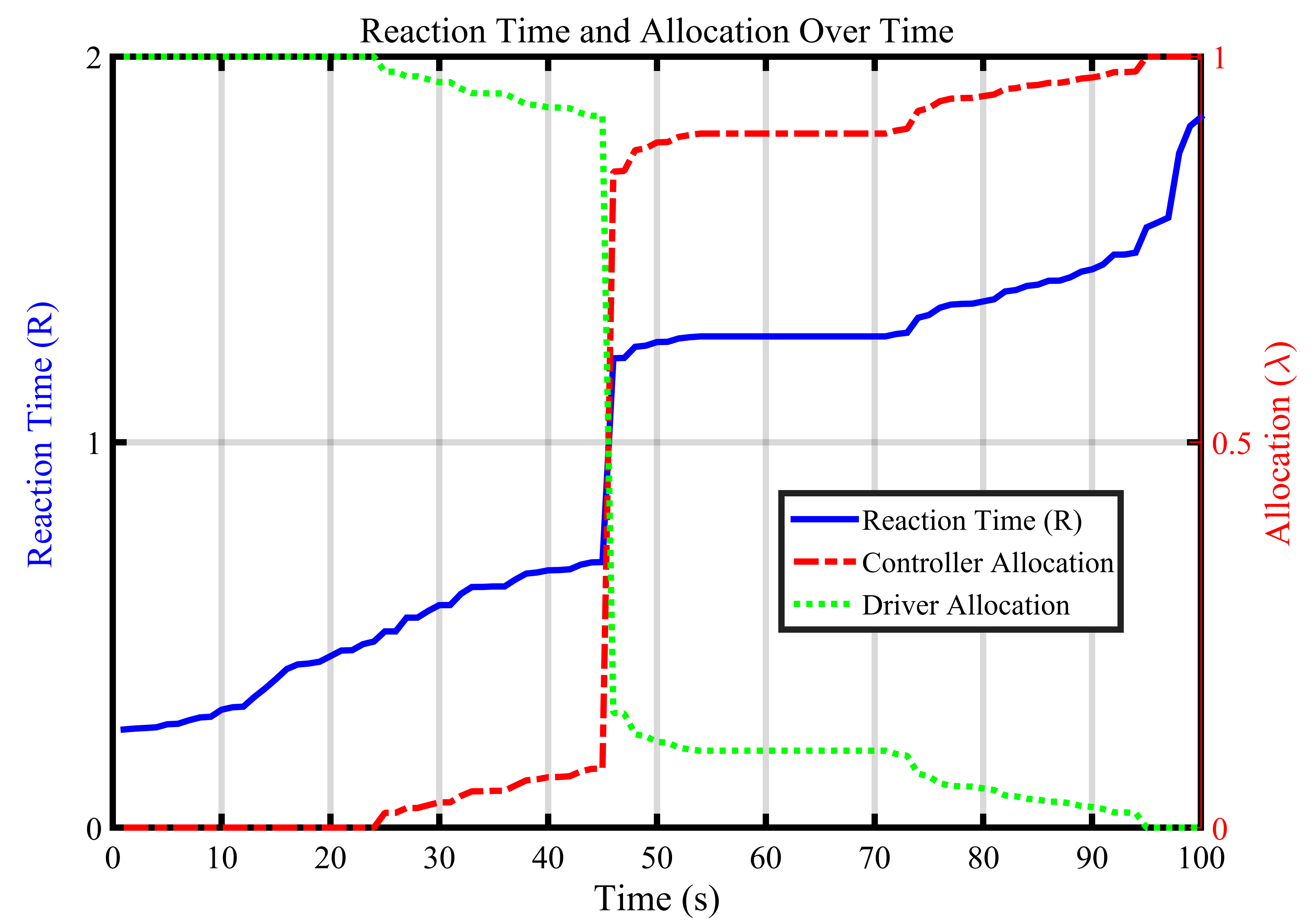}
    \caption{Reaction time and control allocation over time.}
    \label{fig:rac}
\end{figure}
\begin{figure}[!htbp]
    \centering
    \includegraphics[width=3in]{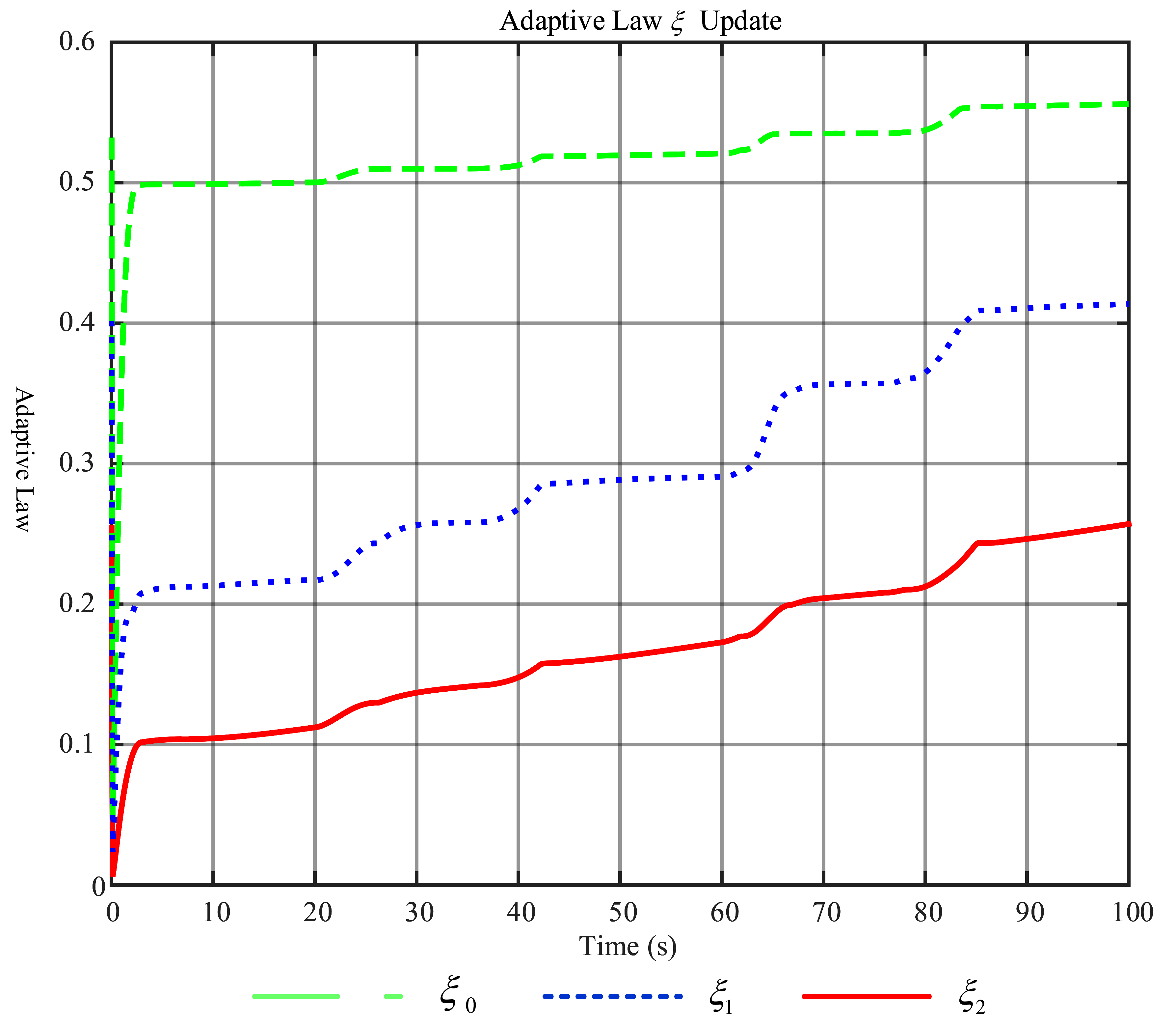}
    \caption{Adaptive law update under sudden changes in driver reaction time.}
    \label{fig:law}
\end{figure}
\begin{figure*}[!htbp]
    \centering
    \includegraphics[width=6.8in]{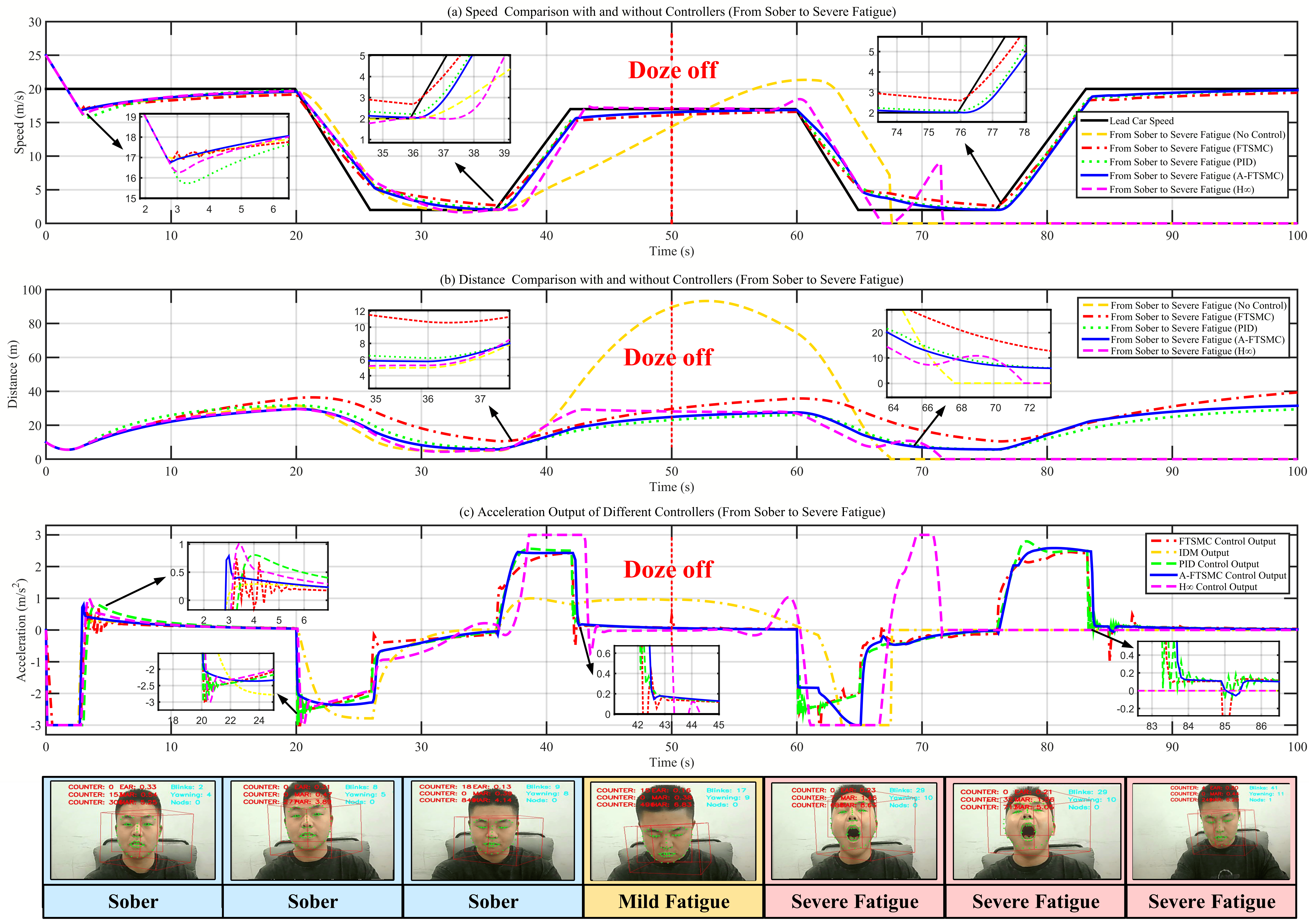}
    \caption{Performance of different controllers under sudden changes in driver reaction time.}
    \label{fig:dsc}
\end{figure*}

Fig. \ref{fig:jgg} demonstrates the comparison of speed, distance, and acceleration under different driver states using various controllers. In terms of velocity tracking, as the reaction time increases, the FTSMC controller exhibits noticeable chattering, the PID controller shows excessive speed variations, and the \(H_\infty\) controller's response accuracy significantly decreases. In contrast, the A-FTSMC controller maintains good speed tracking performance at reaction times of \ensuremath{0.2\!\ \text{s}}, \ensuremath{1.2\!\ \text{s}}, and \ensuremath{2.0\!\ \text{s}}, adjusting quickly to match the lead vehicle's speed.

In terms of distance maintenance, the A-FTSMC controller stabilizes the relative distance at a pre-set safe following distance of around \ensuremath{5\!\ \text{m}}. The PID controller's maximum tracking distance is \ensuremath{1.6\!\ \text{m}}-\ensuremath{2.3\!\ \text{m}} greater than other controllers, affecting road capacity. The FTSMC controller cannot stabilize the distance at the optimal distance but instead stabilizes at around \ensuremath{8.7\!\ \text{m}}. As reaction time increases, the \(H_\infty\) controller fails to maintain safe distance tracking performance, resulting in collisions.

In terms of acceleration, the A-FTSMC controller shows smoother acceleration changes at all reaction times. The maximum acceleration and deceleration remain around \ensuremath{2.6\!\ \text{m/s}^2}, while the PID and the FTSMC controllers have a maximum acceleration of \ensuremath{3\!\ \text{m/s}^2}. When the lead vehicle's speed changes, the PID and the FTSMC controllers take longer (\ensuremath{4.6\!\ \text{s}} or more) to stabilize the acceleration, while the A-FTSMC controller only takes about \ensuremath{2.7\!\ \text{s}}. Additionally, the acceleration jerk amplitude of the PID and the FTSMC controllers is larger (\ensuremath{\pm0.8\!\ \text{m/s}^2}) compared to the A-FTSMC controller (\ensuremath{\pm0.4\!\ \text{m/s}^2}).

Fig. \ref{fig:ada1} illustrates the update of adaptive control law parameters (\(\xi_0\), \(\xi_1\), \(\xi_2\)) under different driver reaction times. As time progresses, each parameter gradually stabilizes. When the reaction time is \ensuremath{0.2\!\ \text{s}}, the parameters converge quickly and remain at lower levels. For reaction times of \ensuremath{1.2\!\ \text{s}} and \ensuremath{2.0\!\ \text{s}}, the parameters converge more slowly and to higher values. This indicates that as the driver reaction time increases, the adaptive controller chooses higher parameter adjustments to compensate for the delays, ensuring system stability.

Fig. \ref{fig:err} compares distance errors under various driver reaction times using different controllers. The A-FTSMC controller keeps the error within \ensuremath{\pm1.8\!\ \text{m}} and stabilizes in \ensuremath{3\!\ \text{s}} at a \ensuremath{0.2\!\ \text{s}} reaction time, while the FTSMC and PID controllers show larger fluctuations, with PID reaching \ensuremath{3.8\!\ \text{m}}. The \(H_{\infty}\) and FTSMC controllers have significant fluctuations, around \ensuremath{8.2\!\ \text{m}} and \ensuremath{9.8\!\ \text{m}}. As reaction time increases to \ensuremath{1.2\!\ \text{s}}, A-FTSMC maintains errors within \ensuremath{\pm10\!\ \text{m}} and stabilizes in \ensuremath{6.2\!\ \text{s}}, while others peak at \ensuremath{20\!\ \text{m}}, taking \ensuremath{8.1\!\ \text{s}} to stabilize. At \ensuremath{2.0\!\ \text{s}}, A-FTSMC holds errors within \ensuremath{\pm20\!\ \text{m}} and stabilizes in \ensuremath{8.4\!\ \text{s}}, with other controllers exceeding \ensuremath{30\!\ \text{m}}. Overall, A-FTSMC improves stability and response time by better handling model uncertainties.

Fig. \ref{fig:acc} compares the acceleration outputs of the A-FTSMC controller and the IDM-RTD model under different driver reaction times. At a reaction time of \ensuremath{0.2\!\ \text{s}}, the IDM-RTD controller shows significant fluctuations but generally follows the lead vehicle's acceleration. The A-FTSMC controller is more stable, with an error within \ensuremath{\pm0.5\!\ \text{m/s}^2}, resulting in better overall performance. As the reaction time increases to \ensuremath{1.2\!\ \text{s}}, IDM-RTD’s fluctuations decrease but lag becomes more apparent. The A-FTSMC controller compensates by increasing output weight, maintaining error within \ensuremath{\pm0.8\!\ \text{m/s}^2}, and despite some lag, the system continues to follow the lead vehicle effectively. At \ensuremath{2.0\!\ \text{s}}, IDM-RTD’s lag worsens, but A-FTSMC improves system response and accuracy, controlling error within \ensuremath{\pm1.1\!\ \text{m/s}^2}. Overall, A-FTSMC takes on more control as reaction time increases, effectively compensating for driver delays and enhancing system robustness.

\subsection{Simulation with Sudden Changes in Driver State}

In real driving environments, a driver's state can change suddenly, causing unpredictable variations in reaction time. This makes it essential to validate the controller's performance under uncertainty scenarios. To study this phenomenon, this work selected key scenarios segments with sudden changes from the pre-recorded video data. The reaction time of the driver in the video was calculated using the fuzzy reaction time evaluation model presented in Section \ref{section2}.

Fig. \ref{fig:rac} shows variations in driver reaction time, controller allocation, and driver allocation. Initially, with a reaction time of about \ensuremath{0.2\!\ \text{s}}, controller intervention is minimal. As the driver becomes fatigued, reaction time increases, spiking to nearly \ensuremath{2\!\ \text{s}} around \ensuremath{40\!\ \text{s}}. This leads to a significant rise in controller allocation to compensate for the delay.

Figs. \ref{fig:law} and \ref{fig:dsc} compare speed, distance, and acceleration under different driver states (sober, mild, and severe fatigue) using various controllers. In the sober state (\ensuremath{0\!\ \text{s}} to \ensuremath{40\!\ \text{s}}), the A-FTSMC controller maintains speed error within \ensuremath{\pm0.9\!\ \text{m/s}}, while PID and \(H_{\infty}\) show larger errors, up to \ensuremath{\pm1.8\!\ \text{m/s}} and \ensuremath{\pm2.9\!\ \text{m/s}}. After the sudden reaction time spike at \ensuremath{50\!\ \text{s}}, A-FTSMC stabilizes within \ensuremath{3.1\!\ \text{s}}, while \(H_{\infty}\) exhibits the largest fluctuations, reaching \ensuremath{\pm7.2\!\ \text{m/s}}.In terms of distance, A-FTSMC maintains error within \ensuremath{\pm5\!\ \text{m}} in the sober state and within \ensuremath{\pm10\!\ \text{m}} after the spike. In contrast, \(H_{\infty}\) shows significant fluctuations, peaking at \ensuremath{20\!\ \text{m}} and resulting in a collision at \ensuremath{73\!\ \text{s}}.For acceleration, A-FTSMC keeps error within \ensuremath{\pm0.5\!\ \text{m/s}^2}, quickly stabilizing after the spike. Other controllers, especially \(H_{\infty}\), show greater fluctuations, with errors reaching \ensuremath{\pm3\!\ \text{m/s}^2} after \ensuremath{60\!\ \text{s}}.

Overall, A-FTSMC demonstrates faster stabilization and smaller errors across all states, with stabilization times reduced by \ensuremath{27.3\%\!} , showing better robustness and adaptability in handling uncertainties.

\section{Conclusion}

This paper built an advanced control framework for the HM-VFAS aimed at mitigating instability caused by time-varying driver states. To capture vehicle-following behavior under different driver conditions, driver state is assessed through facial feature extraction to estimate reaction times, which are integrated into the IDM-RTD model. To manage time-varying driver output, a dynamic control authority allocation strategy between the driver and ADAS is implemented. A two-layer A-FTSMC is then applied to achieve finite-time system stabilization. The first layer compensates for uncertainties in driver states, ensuring robustness, while the second layer accelerates system convergence. Finally, real driver videos are conducted to validate system performance. The results show substantial improvements in driving safety and stability, confirming the effectiveness of the proposed controller.

In the future, this work will include distracted driving and other abnormal behaviors as evaluation criteria for assessing the driver's state. In addition, the shared control system will undergo hardware-in-the-loop (HIL) testing to ensure its robustness and effectiveness in practical implementations.

\appendix\label{tiqu}

The appendix provides a description of the algorithms in Section II(B) for calculating the eye feature vector , mouth feature vector , and motion entropy. The pseudo-code for these calculations is displayed in Algorithm \ref{tq}.

\begin{algorithm}[h]
\small
\caption{Calculation of EFV, MFV, $H_F(X)$}
\label{tq}
\KwIn{$p_i$ (feature points coordinates), $N_{\text{total}}$ (total number of frames), $N_{\text{close}}$ (number of closed-eye frames)}
\KwOut{EFV, MFV, $H_F(X)$}
\BlankLine

\SetKwFunction{CalculateEFV}{Calculate\_EFV}
\SetKwFunction{CalculateMFV}{Calculate\_MFV}
\SetKwFunction{CalculateHF}{Calculate\_HF}

\SetKwProg{Fn}{Function}{:}{}
\Fn{\CalculateEFV{$p_1, p_2, p_3, p_4, p_5, p_6$}}{
    // Calculate Eye Feature Vector (EFV)\;
    // EFV is used to measure the state of the driver's eyes\;
    $EFV \gets \frac{\|p_2 - p_6\| + \|p_3 - p_5\|}{2 \|p_1 - p_4\|}$\;
    \KwRet EFV\;
}

\BlankLine

\Fn{\CalculateMFV{$p_{11}, p_{12}, p_{13}, p_{14}, p_{15}, p_{16}, p_{17}, p_{18}$}}{
    // Calculate Mouth Feature Vector (MFV)\;
    // MFV is used to measure the state of the driver's mouth\;
    $MFV \gets \frac{\|p_{12} - p_{16}\| + \|p_{13} - p_{17}\| + \|p_{14} - p_{18}\|}{3 \|p_{11} - p_{15}\|}$\;
    \KwRet MFV\;
}

\BlankLine

\Fn{\CalculateHF{$p_1, p_2, \ldots, p_N$}}{
    // Calculate Motion Entropy $H_F(X)$\;
    // $H_F(X)$ reflects the randomness of facial motion features\;
    $N \gets \text{number of feature points}$\;
    $F_x \gets \frac{1}{N} \sum_{i=1}^{N} p_{ix}, \quad F_y \gets \frac{1}{N} \sum_{i=1}^{N} p_{iy}$\;
    $l_i \gets \sqrt{(F_x - p_{ix})^2 + (F_y - p_{iy})^2}$ for $i = 1, 2, \ldots, N$\;
    $\mu_l \gets \frac{1}{N} \sum_{i=1}^N l_i$\;
    $\sigma_l \gets \sqrt{\frac{1}{N} \sum_{i=1}^N (l_i - \mu_l)^2}$\;
    // Define intervals to group distances\;
    $I_i \gets \left( (i-1) \frac{\mu_l}{\sigma_l}, i \frac{\mu_l}{\sigma_l} \right)$ for $i = 1, 2, \ldots, i_{\text{max}}$\;
    $i_{\text{max}} \gets \left( \frac{\max(l_1, l_2, \ldots, l_N)}{\mu_l / \sigma_l} + 1 \right)$\;
    $n_i \gets \text{count of } l_i \text{ in each interval } I_i$\;
    // Calculate the probability distribution of distances\;
    $q(l_i) \gets \frac{n_i}{N}$\;
    // Calculate the motion entropy to reflect the variability in facial motion\;
    $H_F(X) \gets - \sum_{i=1}^{i_{\text{max}}} q(l_i) \log q(l_i)$\;
    \KwRet $H_F(X)$\;
}

\BlankLine

EFV $\gets$ \CalculateEFV{$p_1, p_2, p_3, p_4, p_5, p_6$}\;
MFV $\gets$ \CalculateMFV{$p_{11}, p_{12}, p_{13}, p_{14}, p_{15}, p_{16}, p_{17}, p_{18}$}\;
$H_F(X)$ $\gets$ \CalculateHF{$p_1, p_2, \ldots, p_N$}\;
\end{algorithm}

\section*{Acknowledgement}
This project is jointly supported by National Natural Science Foundation of China (Nos. 52172350, 51775565), Guangdong Basic and Applied Research Foundation (Nos. 2021B1515120032, 2022B1515120072), Guangzhou Science and Technology Plan Project (Nos. 2024B01W0079, 202206030005), Nansha Key RD Program (No. 2022ZD014), Science and Technology Planning Project of Guangdong Province (No. 2023B1212060029).
\bibliographystyle{elsarticle-num} 
\bibliography{refrence}
\vspace{0 mm}
\begin{IEEEbiography}[{\includegraphics[width=1in,height=1.25in,clip,keepaspectratio]{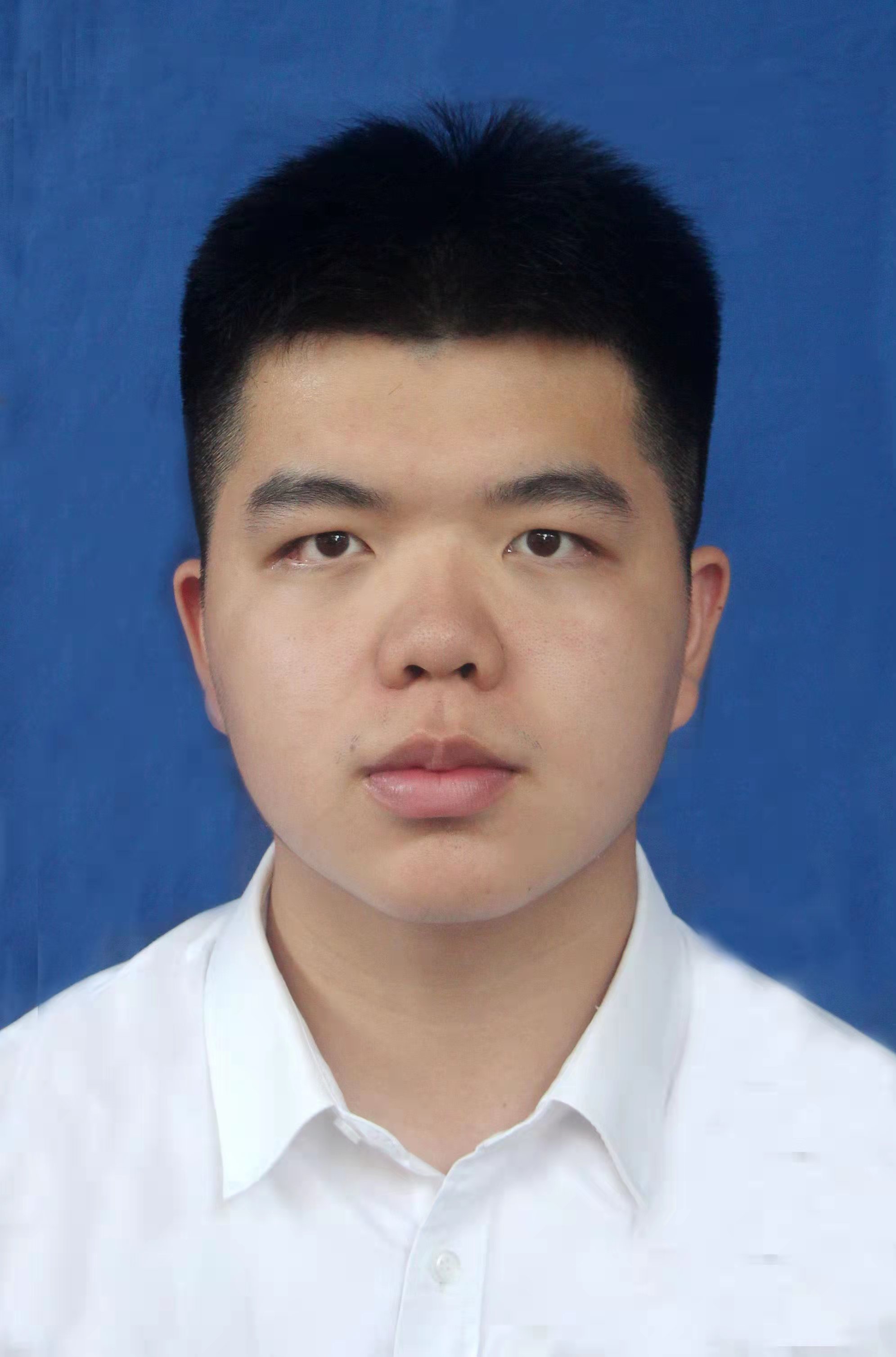}}]{Zihan Wang}
received his B.S. degree in Traffic Engineering from Chang’an University, Xian, China, in 2021. He is currently working towards his master’s degree at Sun Yat-sen University, Guangzhou, China. His current research interests include autonomous driving and vehicle-road collaboration, deep learning and computer vision. 
\end{IEEEbiography}
\vspace{-10 mm}
\begin{IEEEbiography}[{\includegraphics[width=1in,height=1.25in,clip,keepaspectratio]{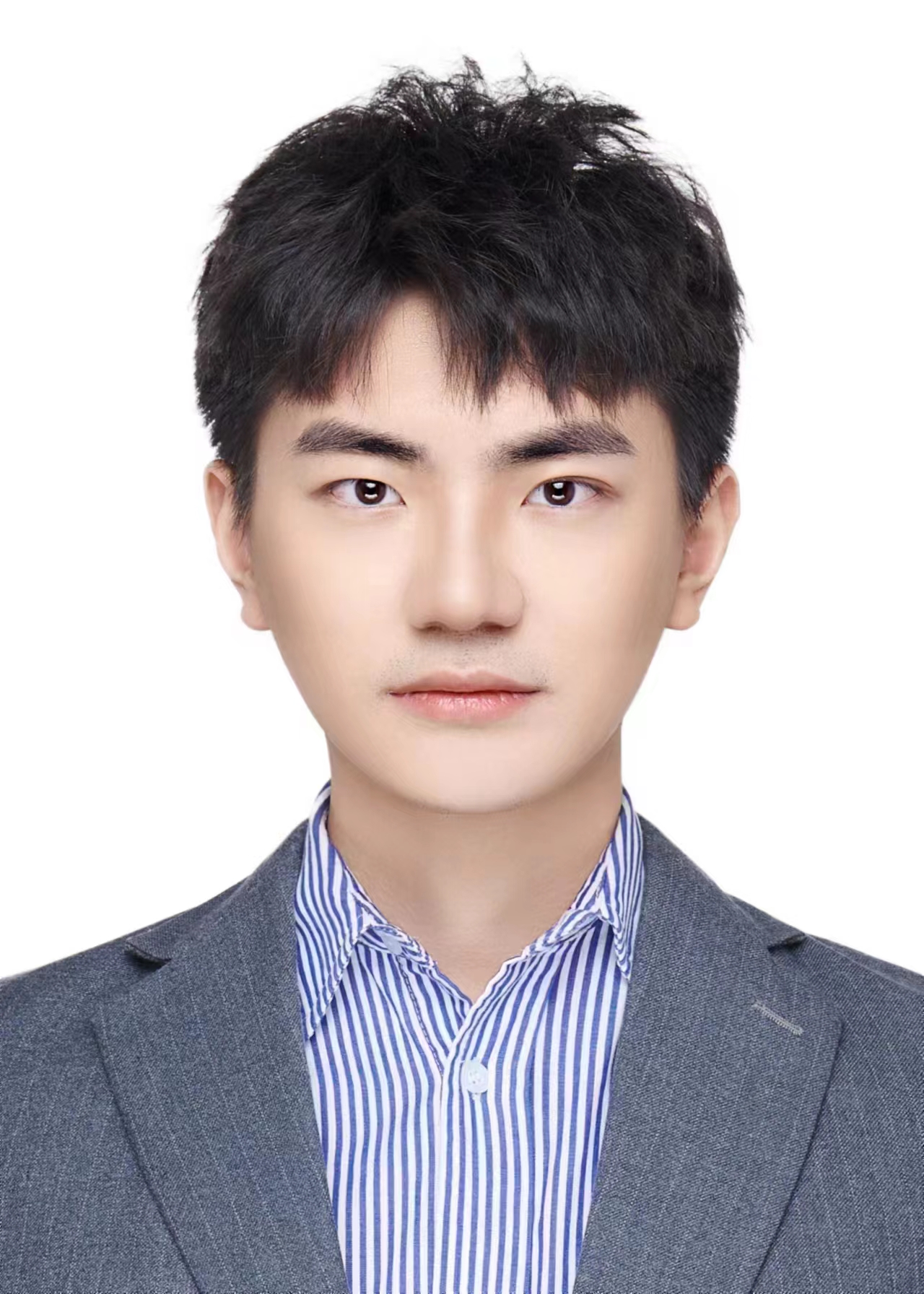}}]
{Mengran Li} is currently pursuing a Ph.D. degree at the Guangdong Key Laboratory of Intelligent Transportation Systems, School of Intelligent Systems Engineering, Sun Yat-sen University, Shenzhen, China. He received an M.S. degree in Control Science and Engineering from the Beijing Key Laboratory of Multimedia and Intelligent Software Technology, Beijing University of Technology, Beijing, China, in 2023. His research interests include big data and artificial intelligence.
\end{IEEEbiography}
\vspace{-10 mm}
\begin{IEEEbiography}
[{\includegraphics[width=1in,height=1.25in,clip,keepaspectratio]{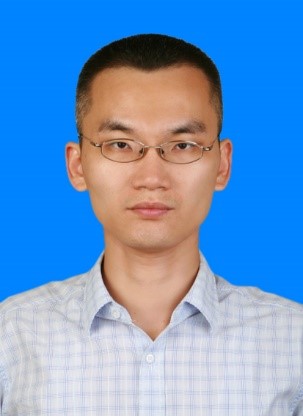}}]{Ronghui Zhang}
received a B.Sc. (Eng.) from the Department of Automation Science and Electrical Engineering, Hebei University, Baoding, China, in 2003, an M.S. degree in Vehicle Application Engineering from Jilin University, Changchun, China, in 2006, and a Ph.D. (Eng.) in Mechanical \& Electrical Engineering from Changchun Institute of Optics, Fine Mechanics and Physics, the Chinese Academy of Sciences, Changchun, China, in 2009. After finishing his post-doctoral research work at INRIA, Paris, France, in February 2011, he is currently an Associate Professor with the Research Center of Intelligent Transportation Systems, School of intelligent systems engineering, Sun Yat-sen University, Guangzhou, Guangdong 510275, P.R.China. His current research interests include computer vision, intelligent control and ITS.
\end{IEEEbiography}
\vspace{-10 mm}
\begin{IEEEbiography}
[{\includegraphics[width=1in,height=1.25in,clip,keepaspectratio]{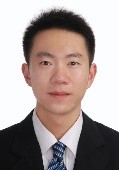}}]{Jing Zhao}
received the Ph.D. degree in electromechanical engineering from University of Macau, Macao. He is currently working with Automotive Engineering Lab, Department of Electromechanical Engineering, University of Macau. His research interests include system dynamics and control, advanced control of intelligent systems.  
\end{IEEEbiography}
\vspace{-10 mm}
\begin{IEEEbiography}
[{\includegraphics[width=1in,height=1.25in,clip,keepaspectratio]{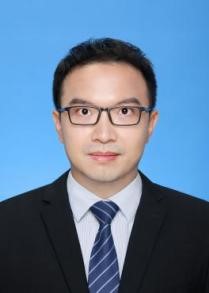}}]{Chuan Hu}
is now a tenure-track Associate Professor at the School of Mechanical Engineering, Shanghai Jiao Tong University, Shanghai, China, starting from July, 2022. Before that, he was an Assistant Professor at the Department of Mechanical Engineering, University of Alaska Fairbanks, Fairbanks, AK, USA from June, 2020 to June, 2022. He was a Postdoctoral Fellow at Department of Mechanical Engineering, University of Texas at Austin, Austin, USA, from August, 2018 to June, 2020, and a Postdoctoral Fellow in the Department of Systems Design Engineering, University of Waterloo, Waterloo, Canada from July, 2017 to July, 2018. He received the Ph.D. degree in Mechanical Engineering, McMaster University, Hamilton, Canada in 2017, the M.S. degree in Vehicle Operation Engineering from the China Academy of Railway Sciences, Beijing, in 2013, and the B.S. degree in Automotive Engineering from Tsinghua University, Beijing, China, in 2010. His research interest includes the perception, decision-making, path planning, and motion control of Intelligent and Connected Vehicles (ICVs), Autonomous Driving (AD), eco-driving, human-machine trust and cooperation, shared control, and machine-learning applications in ICVs. He has published more than 60 papers in these research areas. He is currently an Associate Editor for several leading IEEE Trans. journals in this filed: IEEE Trans. on Neural Networks and Learning Systems, IEEE Trans. on Vehicular Technology, IEEE Trans. on Transportation Electrification, IEEE Trans. on Intelligent Transportation Systems and IEEE Trans. on Intelligent Vehicles. 
\end{IEEEbiography}
\vspace{-10 mm}
\begin{IEEEbiography}[{\includegraphics[width=1in,height=1.10in,clip,keepaspectratio]{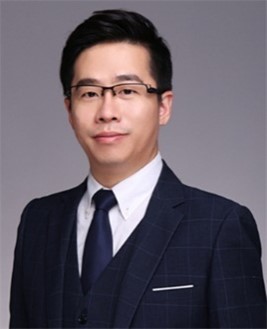}}]{Xiaolei Ma} (Senior Member, IEEE) earned his Ph.D. degree from the University of Washington, Seattle, WA, USA, in 2013. He currently serves as a Professor at the School of Transportation Science and Engineering, Beihang University, China. His research focuses on public transit operations and planning, transportation big data analytics, and the integration of transportation and energy. Dr. Ma is an associate editor for the IEEE Transactions on Intelligent Transportation Systems and IET Intelligent Transport Systems. He also contributes as an editorial member to several journals, including Transportation Research Part C and D, Computers, Environment and Urban Systems, and the Journal of Intelligent Transportation Systems. Dr. Ma is a Fellow of the IET and a Senior Member of the IEEE.
\end{IEEEbiography}
\vspace{-10 mm}
\begin{IEEEbiography}[{\includegraphics[width=1in,height=1.25in,clip,keepaspectratio]{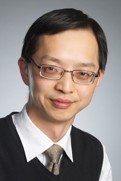}}]{Zhijun Qiu}
is a Professor in the Faculty of Engineering at University of Alberta, Canada Research Chair Professor in Cooperative Transportation Systems, and Director of Centre for Smart Transportation. His research interest includes traffic operation and control, traffic flow theory, and traffic model analytics. He has published more than 180 papers in international journals and academic conferences, and has 7 awarded patents and 5 pending application patents. Dr. Qiu is working as the Managing Director for the ACTIVE-AURORA test bed network, which is Canadian National Connected Vehicle Test Bed, and sponsored by Transport Canada, City of Edmonton, Alberta Transportation and other funding agencies, to identify leading-edge Connected Vehicle solutions through research and development. His theoretical research has informed and enriched his many practical contributions, which have been widely supported by private and public sectors. Dr. Tony Qiu received his PhD degree from University of Wisconsin-Madison, and worked as a Post-Doctoral Researcher in the California PATH Program at the University of California, Berkeley before joining University of Alberta. Dr. Tony Qiu has been awarded Minister’s Award of Excellence in 2013, Faculty of Engineering Annual Research Award in 2015-2016, and ITS Canada Annual Innovation and R$\And$D Award in 2016 and 2017.
\end{IEEEbiography}
\end{document}